\documentclass[11pt]{iopart}
\usepackage{bm}
\usepackage{hyperref}
\usepackage{amssymb}
\usepackage{feynmp}
\usepackage{amsopn}
\usepackage{amsbsy}
\usepackage{setstack}
\usepackage{deltandiagrams}

\newcommand{\ie}{\emph{i.e.}}
\newcommand{\lp}{\mathrm{1loop}}
\renewcommand{\etal}{\emph{et al.}}

\newcommand{\R}{\mathcal{R}}

\newcommand{\Ps}{\mathcal{P}}
\newcommand{\Planck}{M_{\mathrm{P}}}
\newcommand{\z}{\bm{\mathrm{z}}}
\newcommand{\flabel}[1]{{\small(\textit{{#1}})}}

\newcommand{\transpose}{\mathsf{T}}
\newcommand{\dfin}[1]{\delta_{{#1}\fin}}
\newcommand{\fin}{\sharp}
\newcommand{\Sghost}{S_{\mathrm{gh}}}

\renewcommand{\d}{\mathrm{d}}
\newcommand{\vect}[1]{\bm{\mathrm{{#1}}}}
\renewcommand{\e}[1]{\mathrm{e}^{{#1}}}

\newcommand{\im}{\mathrm{i}}

\newcommand{\grad}{\nabla}

\renewcommand{\geq}{\geqslant}
\newcommand{\laplacian}{\triangle}
\DeclareMathOperator{\ord}{o}
\DeclareMathOperator{\ImPart}{Im}

\renewcommand{\Im}{\ImPart}

\newenvironment{equation-diagram}{\vspace{3mm}\begin{equation}}
  {\end{equation}\par\vspace{-3mm}\noindent\ignorespaces\hspace{-0.7ex}}
\newenvironment{eqnarray-diagram}{\vspace{3mm}\begin{eqnarray}}
  {\end{eqnarray}\par\vspace{-3mm}\noindent\ignorespaces\hspace{-0.7ex}}

\makeatletter
\newcommand\numberwithin[2]{\@addtoreset{#1}{#2}}
\makeatother
\numberwithin{footnote}{section}

\begin{document}
\begin{fmffile}{diags}
	\title{One-loop corrections to a scalar field during inflation}
	\date{\today}
	\author{David Seery}
	\address{Astronomy Unit, School of Mathematical Sciences\\
	  Queen Mary, University of London\\
	  Mile End Road, London E1 4NS\\
	  United Kingdom}
	\eads{\mailto{D.Seery@qmul.ac.uk}}
	\submitto{JCAP}
	\begin{abstract}
		The leading quantum correction to the power spectrum of a
		gravitationally-coupled
		light scalar field is calculated, assuming that it is
		generated during a phase of single-field, slow-roll inflation.
	\vspace{3mm}
	\begin{flushleft}
	\end{flushleft}
	\end{abstract}
	\maketitle

	\section{Introduction}
	Over the last decade, our theories of the early universe have
	been promoted from speculation to a field of intense
	scientific study. The most important developments in our knowledge
	concern the nature of the primordial curvature perturbation $\zeta$%
  		\footnote{There are two primordial perturbations commonly encountered
		in the literature.
		The first of these is the \emph{comoving curvature
		perturbation}, written $\R$, which is proportional to the laplacian
		of the Ricci curvature of comoving spatial slices. On the other
	 	hand, the \emph{uniform density curvature perturbation} $\zeta$ is
		proportional to the laplacian of the Ricci curvature on spatial
		slices of uniform density. On superhorizon scales,
		$\R$ and $\zeta$ are equivalent up to a convention for signs
		\cite{Wands:2000dp}.},
	which is believed to have seeded temperature variations in the
	cosmic microwave background (CMB). It is now understood that $\zeta$
	must have had a spectrum which was close to scale invariance on the
	scales probed by the CMB \cite{Spergel:2006hy,Martin:2006rs,Kinney:2006qm}.
	
	Many proposals have been made to explain how a primordial
	perturbation with an almost scale-invariant spectrum could have
	been generated in the early universe. The most widely-studied
	candidate is an era of inflation that may have
	taken place at high energy
	\cite{Starobinsky:1980te,Sato:1980yn,Guth:1980zm,
	Hawking:1981fz,Albrecht:1982wi,Linde:1981mu,Linde:1983gd},
	where ``inflation'' is defined to be any epoch in which the scale
	factor $a$ undergoes acceleration, $\ddot{a} > 0$. Under this condition,
	local regions of the universe are exponentially driven to
	spatial flatness, homogeneity and isotropy
	\cite{Wald:1983ky}, and each light bosonic field acquires a
	perturbation generated by amplification of
	quantum-mechanical vacuum fluctuations
	\cite{Bardeen:1983qw,Guth:1982ec,Hawking:1982cz,Hawking:1982my}.
	The spectrum of this perturbation is close
	to scale-invariance when the universe inflates at a rate
	$\dot{a}/a$ which is almost constant. The curvature
	perturbation observed in the CMB is supposed to be a model-dependent
	mix of these fluctuations, yielding anisotropies in the
	temperature of the microwave sky which are compatible with observation.
	Inflation apparently provides a natural framework
	in which one can simultaneously understand both the
	large-scale regularity of the universe \emph{and} its small-scale
	irregularity.
	
	Inflation is not a single model, but rather a whole collection
	of scenarios which fit into the above framework. The only
	necessary ingredients are: (i) a specification of the field
	content, which allows a division into `light' and `heavy'
	fields; (ii) a background evolution $a(t)$ which gives rise to
	$\ddot{a} > 0$ with the Hubble parameter $H \equiv \dot{a}/a$
	slowly varying;
	and (iii) a rule for generating $\zeta$ from the light
	bosonic fields.
	
	This prescription is rather general and implies that
	many models (perhaps with wildly different and mutually
	incompatible microphysics) may simultaneously be compatible
	with the observational data, since they may make equivalent
	predictions for the spectrum of $\zeta$.
	Therefore, we must expect that
	it will be difficult to learn about the
	microscopic physics which was operative during the very early
	universe: it will almost certainly be insufficient
	simply to study the spectrum of $\zeta$. In order to distinguish
	between wildly different models of the early universe it
	is necessary to find another source of information.
	
	Fortunately, any detailed model of the inflationary era does not merely 
	predict the spectrum of $\zeta$; it also implies a subtle but
	calculable network of correlations between the higher-order
	moments. These moments collectively measure the so-called
	non-gaussianity of $\zeta$ and arise from interactions
	among the quanta of the $\zeta$-field and the other constituents of the
	early universe.
	Non-gaussian effects from interactions of $\zeta$ quanta
	have been extensively investigated over the last few
	years, with the hope that observations of such effects may
	be able to discriminate between different models for physics
	in the early universe
	\cite{Bartolo:2004if,Maldacena:2002vr,Rigopoulos:2004gr,Rigopoulos:2004ba,
	Seery:2005gb,Rigopoulos:2005xx,Lyth:2005fi,Lyth:2005qj,Zaballa:2006pv,
	Battefeld:2006sz,Battefeld:2007en,Kim:2006te,
	Allen:2005ye,Alabidi:2006hg,Gupta:2002kn,Gupta:2005nh,
	Lyth:2002my,
	Boubekeur:2005fj,Lyth:2005du,Lyth:2006gd,Malik:2006pm,Enqvist:2005pg,
	Enqvist:2004ey,Huang:2005nd,Barnaby:2006cq,Barnaby:2007yb,
	Valiviita:2006mz,Byrnes:2006vq,Byrnes:2007tm}.
	However, more is possible.
	Interactions do not only imply non-gaussian statistics
	in the three- and higher $n$-point correlation functions:
	they also imply \emph{quantum} corrections to all
	correlation functions, and in particular the power spectrum
	or two-point function. It is possible that such corrections
	may be large in their own right, demanding that they be taken
	into account in accurate analyses of the observational data,
	as recently suggested by Sloth
	\cite{Sloth:2006az,Sloth:2006nu}.
	Regardless of their exact magnitude,
	by searching for signatures of such
	quantum corrections in the power spectrum and
	correlating the results with predictions for non-gaussian
	statistics in the higher $n$-point functions we obtain
	a more sensitive test of physics during inflation.
	
	A second powerful motivation for studying loop corrections
	is a simple point of principle. The tree-level formula for the
	spectrum of $\zeta$ is widely used to make predictions for
	the amplitude and scale-dependence of fluctuations generated
	in a very large class of early universe scenarios. Before
	deciding what degree of credence we should attach to any of
	these predictions, it is necessary to thoroughly investigate
	whether the tree-level amplitude is a genuine approximation
	to the full quantum result.
	
	In this paper, the prospects for detecting quantum corrections
	to the power spectrum are assessed in the simplest model of
	inflationary physics, that of inflation with a single scalar
	$\phi$ and arbitrary potential $V(\phi)$. In view of the
	importance of accurate comparison with the precision
	measurements which are
	becoming available, this issue has already attracted
	considerable attention.
	Early work by Mukhanov, Abramo \& Brandenberger
	\cite{Mukhanov:1996ak,Abramo:1997hu}
	and Abramo \& Woodard
	\cite{Abramo:1998hi} demonstrated that significant
	effects were possible
	(see also Unruh \cite{Unruh:1998ic}).
	Later estimates of loop effects were made in a large number of models
	\cite{Prokopec:2002uw,Brandenberger:2002sk,Onemli:2002hr,Prokopec:2003bx,
	Geshnizjani:2003cn,Brandenberger:2004ki,
	Brandenberger:2004kx,Brandenberger:2004ix,
	Onemli:2004mb,Brunier:2004sb,
	Boyanovsky:2005sh,Boyanovsky:2005px,Kahya:2005kj,Weinberg:2005vy,
	Martineau:2005aa,
	Chaicherdsakul:2006ui,Losic:2005vg,Kahya:2006hc,Bilandzic:2007nb,
	Kahya:2006ui,Kim:2007sb}, even at two-loop order \cite{Prokopec:2006ue}.
	Recently, Sloth	\cite{Sloth:2006az,Sloth:2006nu} determined the
	full fourth-order action for Einstein gravity coupled to a scalar field
	and used this to estimate the one-loop correction to the power
	spectrum of scalar field fluctuations.
	Although one would na\"{\i}vely expect the loop correction to be
	suppressed by a factor of $(H/\Planck)^2 \sim 10^{-10}$,
	where $H$ is the Hubble parameter during inflation,
	Sloth's calculation yielded a significant cumulative effect
	(as large as 70\% in some models)
	which could affect the precision determination of cosmological parameters
	from CMB experiments.
	The unexpectedly large size of the loop correction in this estimate is
	due to an amplification by $N$, the total number of e-folds
	of inflation which occur. Since $N$ can be very large in models
	where inflation begins at around the Planck scale, it may dramatically
	modify the predictions of na\"{\i}ve dimensional analysis.
	
	This paper attempts to address these issues using a
	formalism similar
	to that applied by Sloth \cite{Sloth:2006az,Sloth:2006nu}.
	However, in contrast to previous analyses, the
	estimate is divided into two parts. First, the one-loop
	correction to the power spectrum of field fluctuations is
	computed soon after horizon exit, using the slow-roll approximation
	to control the calculation. This correction is not
	observable by itself; in a second step, it must be combined with
	other correlators of the fields to yield
	the one-loop correction to the power spectrum of the
	observable perturbation $\zeta$ long after horizon exit.
	The correct combination can be computed using the
	$\delta N$ formula
	\cite{Starobinsky:1986fx,Sasaki:1995aw,Rigopoulos:2004gr,
	Lyth:2004gb,Lyth:2005fi}.
	This two-step process has several advantages.
	We shall see that the loop correction is generally afflicted
	by divergences at late times and on large scales.
	The $\delta N$ formalism naturally resums these
	late-time divergences into time evolution, which allows
	the slow-roll approximation to be kept under control.
	On the other hand, the divergences on large
	scales can be controlled by performing the calculation within
	a finite box. In analogy with the late-time divergences,
	it has recently been shown by
	Byrnes {\etal} \cite{Byrnes:2007tm}
	that these divergences can be resummed
	into spatial variation on large scales.
		
	The present paper is concerned with the technical issue
	of computing loop corrections go the power spectrum of
	field fluctuations. This calculation
	involves the application of standard methods from
	quantum field theory, adapted to the case of an expanding
	spacetime.
	On the other hand, the assembly of field correlators
	into $\zeta$ correlators is an essentially classical
	calculation using the $\delta N$ formula. For clarity, this
	calculation will be presented separately elsewhere
	\cite{Seery:2007wf}.
	
	In \S\ref{sec:inflation} the background evolution and
	perturbation theory of the single scalar field are
	briefly described. The perturbations are characterized (as
	in more complex cases) by cubic and higher self-interactions which
	involve the time derivative of the perturbation.
	This has important consequences for the calculation of
	quantum corrections. These corrections are introduced in
	\S\ref{sec:quantum}. In \S\ref{sec:quantum-oneloop} a
	path-integral expression for a general one-loop, single-vertex
	correction to the
	power spectrum is given in the Schwinger formalism, and in
	\S\ref{sec:quantum-derivatives} the question of deriving
	a correct path integral expression for theories with
	time-derivative interactions is considered. In such cases
	the correct path-integral formula is well-known to contain
	a \emph{ghost} field, whose quanta do not appear in physical
	states but circulate in the loops which give rise to
	quantum corrections. The Feynman rules for this theory are
	written down in \S\ref{sec:quantum-rules}.
	In \S\ref{sec:one-point} the assembled formalism is used
	to compute the leading radiative correction to the
	one-point function of the field. This is of interest
	in its own right, but also provides a simple setting in
	which some subtle features of the calculational
	machinery can be resolved. The one-loop correction
	to the two-point function
	is computed in \S\ref{sec:two-point}, and a brief discussion
	is given in \S\ref{sec:discussion}.
	
	\S\ref{sec:inflation} is introductory and merely serves to
	fix notation. The reader who is mostly interested in the
	computation of the two-point function $\langle
	\delta\phi(\vect{k}_1) \delta\phi(\vect{k}_2) \rangle$ may
	wish to skip directly to \S\ref{sec:two-point}
	and dispense with
	\S\S\ref{sec:quantum}--\ref{sec:one-point}.
	These sections are largely
	dominated by the question of setting up a correct
	formalism in which the one-loop correction may be computed.
	
	Units are chosen throughout such that $\hbar = c = \Planck = 1$,
	where $\Planck^{-2} \equiv 8\pi G$ is the reduced Planck mass.
	The metric convention is $(-,+,+,+)$, and the unperturbed
	background is written in cosmic time $t$ as
	\begin{equation}
		\d s^2 = - \d t^2 + a^2(t) \, \d \vect{x}^2 .
		\label{eq:metric}
	\end{equation}
	It is frequently more convenient to employ a conformal
	time variable, defined by $\eta = \int_{\infty}^0
	dt'/a(t')$. Indices labelling spacetime coordinates are
	chosen from the beginning of the Latin alphabet
	$(a, b, \ldots)$; indices labelling purely spatial 
	coordinates are chosen from the middle of the alphabet
	$(i, j, \ldots)$.
	Where multi-field models are under discussion,
	the different species of light bosonic
	fields are labelled with Greek indices $(\alpha, \beta,
	\ldots)$.
	
	\section{Inflation from a single scalar field}
	\label{sec:inflation}
	\subsection{The background evolution}
	
	The simplest microphysical model capable of supporting
	an inflationary epoch consists of Einstein gravity coupled to a
	single scalar field $\phi$ with potential $V(\phi)$, which can
	be taken to be arbitrary except that it must allow
	inflation for some values of $\phi$. The field $\phi$
	is known as the inflaton. The combined action for this system is
	\begin{equation}
		S = - \frac{1}{2} \int \d^4 x \; \sqrt{-g}
		\Big\{ R - \grad^a \phi \grad_a \phi - 2 V(\phi)
		\Big\} ,
		\label{eq:action}
	\end{equation}
	where $g \equiv \det g_{ab}$ is the metric determinant,
	$\d^4 x \equiv \d t \, \d^3 x$ is the product of the spacetime coordinate
	differentials, and $R$ is the
	spacetime Ricci curvature. The background field $\phi$ is taken
	to be spatially homogeneous and the background metric is
	parametrized by the scale factor $a$, given in \eref{eq:metric}.
	The evolution of $a$ is determined by the Friedmann equation
	\begin{equation}
		3 H^2 = \frac{1}{2} \dot{\phi}^2 + V(\phi) ,
		\label{eq:friedmann}
	\end{equation}
	where $H \equiv \dot{a}/a$ is the Hubble parameter, and
	$\phi$ obeys the homogeneous Klein--Gordon equation
	\begin{equation}
		\ddot{\phi} + 3 H \dot{\phi} + V'(\phi) = 0 ,
		\label{eq:kleingordon}
	\end{equation}
	in which a prime $'$ denotes a derivative with respect to
	$\phi$. The condition that inflation occurs is $\ddot{a} > 0$,
	or $\epsilon < 1$, where
	the parameter $\epsilon$ is defined by
	\begin{equation}
		\epsilon \equiv - \frac{\dot{H}}{H^2} .
		\label{eq:epsilon-def}
	\end{equation}
	Using Eqs.~\eref{eq:friedmann}--\eref{eq:kleingordon} one can
	show that an equivalent definition is $\epsilon \equiv \dot{\phi}^2/
	2H^2$.
	
	When $\epsilon$ obeys the stronger condition $\epsilon \ll 1$, the
	rate of change of $\phi$ is negligible in comparison with the
	expansion rate $H$. In this case one says that the field is
	slowly rolling. Although slow-roll is not mandatory for
	inflation to occur, the near scale-invariance of the power spectrum
	imprinted on the scales which are observed in microwave
	background experiments suggests that
	slow-roll was approximately satisfied if the CMB perturbation has an
	inflationary origin. When $\epsilon \ll 1$ applies,
	perturbation theory in $\epsilon$ and related small quantities
	is known as the slow-roll
	approximation. In this paper, we compute all effects to leading order
	in $\epsilon$.
	
	\subsection{Scalar perturbations}
	
	Now consider small spatially-dependent perturbations
	in the inflaton, $\phi = \phi_0 + \delta\phi(t,\vect{x})$, where
	$\phi_0$ is the homogeneous background evolution and $\delta\phi$
	obeys the smallness condition $|\delta\phi| \ll |\phi_0|$.
	Since $\phi$ dominates the energy density of the universe by assumption,
	any perturbation in $\phi$ will lead to a perturbation in the metric.
	These perturbations can be parametrized by a scalar $N$ (the lapse),
	a spatial vector $N^i$ (the shift), and a spatial metric
	$h_{ij}$,
	\begin{equation}
		\d s^2 = - N^2 \, \d t^2 + h_{ij} ( \d x^i + N^i \, \d t )
		( \d x^j + N^j \d t ) .
		\label{eq:adm}
	\end{equation}
	Because of general coordinate invariance, not all choices of
	$\{ N, N^i, h_{ij} \}$ lead to different configurations of the
	gravitational field. This redundancy is removed by fixing a
	gauge.
	We will choose to work in the spatially flat
	gauge, where $h_{ij}$ is given by its background value
	$h_{ij} = a^2(t)\delta_{ij}$. Having done so, the metric functions
	$N$ and $N^i$ are completely determined in terms of $\delta\phi$ by
	the constraints implicit in the Einstein equations.
	
	These can be obtained by inserting
	Eq.~\eref{eq:adm} in the action \eref{eq:action}.
	One obtains
	\begin{equation}
		\fl
		S = - \frac{1}{2} \int \d t \, \d^3 x \; \sqrt{h}
		\left\{ N ( \grad^i \phi \grad_i \phi + 2V) -
		        \frac{1}{N} ( E^{ij} E_{ij} - E^2 + \pi^2 ) \right\} ,
		\label{eq:admaction}
	\end{equation}
	where $E_{ij} = \frac{1}{2} \dot{h}_{ij} - \grad_{(i} N_{j)}$ is the
	``momentum'' associated with $h_{ij}$, $\grad_i$ is the
	spatial covariant derivative compatible with $h_{ij}$, and
	$\pi = \dot{\phi} - N^j \grad_j \phi$ is the field momentum.%
		\footnote{We are adopting a convention, used throughout this
		paper, in which repeated spatial indices in
		complementary raised and lowered positions are contracted with
		the spatial metric $h_{ij}$, whereas a pair of repeated indices
		which both appear in the lowered position are contracted with
		the Euclidean metric $\delta_{ij}$. Thus, $a^i b_i = \sum_{i, j}
		h^{ij} a_i b_j$, whereas $a_i b_i = \sum_i a_i b_i$.
		Spacetime indices obey the usual Einstein convention,
		and always appear in complementary raised and
		lowered pairs which are contracted with the spacetime metric
		$g_{ab}$.}
	The equations of motion for the lapse and shift follow by
	varying $S$ with respect to $N$ and $N^i$ respectively, and do
	not involve time derivatives. Therefore they are not evolution
	equations but constraints and can be solved algebraically:
	$N$ and $N^i$ are not propagating fields. Once $N$ and $N^i$
	are known they may be substituted in \eref{eq:admaction} to
	obtain a reduced action which depends only on $\delta\phi$.
	
	The $N$ constraint is
	\begin{equation}
		\grad^i \phi \grad_i \phi + 2V +
		\frac{1}{N^2}(E^{ij} E_{ij} - E^2 + \pi^2) = 0
		\label{eq:lapse}
	\end{equation}
	and the $N^i$ constraint is
	\begin{equation}
		\grad_i \left\{ \frac{1}{N}(E^i_j - E \delta^i_j) \right\}
		= \frac{\pi}{N} \grad_j \phi .
		\label{eq:shift}
	\end{equation}
	One solves Eqs.~\eref{eq:lapse}--\eref{eq:shift} order by order
	in $\delta\phi$. We write
	\begin{equation}
		N = 1 + \sum_{m=1}^{\infty} \alpha_m,
		\quad \mbox{and} \quad
		N_i = \grad_i \left( \sum_{m=1}^{\infty} \vartheta_m \right)
		+ \sum_{m=1}^{\infty} \beta_{mi}
	\end{equation}
	where $\alpha_m$, $\vartheta_m$ and $\beta_{mi}$ are all $m$th order
	in $\delta\phi$ and the $\beta_{mi}$ are chosen to be divergenceless,
	so that $\partial_i \beta_{mi}$ for all $m$. The expressions necessary
	to compute $S$ to third order in $\delta\phi$ were given by
	Maldacena in the comoving slicing \cite{Maldacena:2002vr} and
	rewritten in the flat slicing for multiple fields in
	Ref.~\cite{Seery:2005gb}.
	The expressions necessary to compute
	$S$ to fourth order were obtained in the flat slicing by
	Sloth \cite{Sloth:2006az,Sloth:2006nu} in an approximation where
	all vector modes were absent, and given in generality
	in Ref.~\cite{Seery:2006vu}.
	
	We work to leading order in the slow-roll approximation. At first
	order in $\delta\phi$ the leading terms are $\ord(\epsilon^{1/2})$ ,
	\begin{equation}
		\alpha_1 = \frac{1}{2H} \dot{\phi} \delta\phi, \quad
		\partial^2 \vartheta_1 = - \frac{a^2}{2H} \dot{\phi}
		\delta{\dot{\phi}}
		\quad \mbox{and} \quad
		\beta_{1i} = 0 .
		\label{eq:scalarvectorone}
	\end{equation}
	At second order in $\delta\phi$ the leading terms are
	$\ord(\epsilon^0)$
	\begin{equation}
		\alpha_2 = \frac{1}{2H} \partial^{-2} \Sigma, \quad
		\frac{4H}{a^2} \partial^2 \vartheta_2 =
		  - \frac{1}{a^2} \partial_i \delta\phi \partial_i \delta\phi
		  - \delta\dot{\phi}\delta\dot{\phi} - 12 H^2 \alpha_2 ,
		\label{eq:scalartwo}
	\end{equation}
	\begin{equation}
		\frac{1}{2a^2} \partial^4 \beta_{2i} = \delta^{rs}
		  ( \partial_i \Sigma_{rs} - \partial_{(r}\Sigma_{s)i} )
		,
		\label{eq:vectortwo}
	\end{equation}
	where bracketed indices $(\cdots)$ are symmetrized with
	total weight unity
	and $\Sigma_{rs}$ is defined by
	\begin{equation}
		\Sigma_{rs} \equiv \partial_r \delta\dot{\phi} \partial_s\delta\phi
		  + \delta\dot{\phi}\partial_r \partial_s \delta\phi ,
		\label{eq:sigma}
	\end{equation}
	with $\Sigma = \tr \Sigma_{rs}$ its trace in the Euclidean metric.
	Eqs.~\eref{eq:scalarvectorone}--\eref{eq:vectortwo} can be inserted
	into the action, Eq.~\eref{eq:admaction}, after which one obtains
	an expansion of $S$ in powers of $\delta\phi$. The first
	non-trivial term is quadratic. At $\ord(\epsilon^0)$ it is equal to
	\begin{equation}
		S_2 = \frac{1}{2} \int \d t \, \d^3 x \; a^3 \left\{
		  \delta\dot{\phi}^2 - \frac{1}{a^2} (\partial\delta\phi)^2
		\right\}
		\label{eq:kinetic}
	\end{equation}
	There is a cubic interaction whose leading term enters at
	$\ord(\epsilon^{1/2})$ \cite{Seery:2005gb}, which can be written
	\begin{equation}
		\fl
		S_3 = \int \d t \, \d^3 x \; a^3 \frac{\dot{\phi}}{4H} \left\{
		  2 \delta\dot{\phi}
		  \partial_j \partial^{-2}
		  \delta\dot{\phi} \partial_j \delta\phi -
		  \delta\phi \left[
		    \delta\dot{\phi}^2 + \frac{1}{a^2}
		    (\partial \delta\phi)^2 \right]
		\right\} .
		\label{eq:cubic}
	\end{equation}
	The quartic term has leading terms of order
	$\ord(\epsilon^0)$ \cite{Seery:2006vu}.
	These terms correspond to
	\begin{equation}
		\fl
		S_4 = \int \d t \, \d^3 x \; \left\{
		  - \frac{1}{4a} \beta_{2j} \partial^2 \beta_{2j} -
		  \frac{a^3}{4H} \partial^{-2} \Sigma \left[
		    \delta\dot{\phi}^2 + \frac{1}{a^2}
		    (\partial\delta\phi)^2 \right]
		  - \frac{3}{4} a^3 (\partial^{-2}\Sigma)^2 -
		  a \delta\dot{\phi}\beta_{2j}\partial_j \delta\phi
		\right\}
		\label{eq:quartic} .
	\end{equation}
	The free field action $S_2$ and the interactions $\{ S_3, S_4 \}$
	as written here are all accompanied by terms of higher-order in slow-roll
	parameters, which we neglect. One must be careful to ensure that this
	approximation is accurate, and we will return to this at various points
	in the analysis (see also Ref.~\cite{Seery:2007wf}).
	
	\subsection{Expectation values}
	\label{sec:expectation-values}
	The observables in this theory are expectation values of products
	of $n$ factors of the perturbation $\delta\phi$, taken at a common
	time $t_{\ast}$ but at distinct spatial coordinates
	$\{ \vect{x}_1, \ldots, \vect{x}_n \}$. It is often more
	convenient to work with momentum space expectation values which
	are obtained by taking Fourier transforms with respect to the
	$\vect{x}_i$, giving $\vect{k}$-space correlators which are
	functions of $\{ \vect{k}_1, \ldots, \vect{k}_n \}$.
	
	At tree level the one-point expectation value vanishes,
	$\langle \delta \phi(\vect{k}) \rangle = 0$, since
	$\delta \phi$ is by definition a perturbation in the comoving
	region under consideration. However, the gravitational background
	is time dependent since the scale factor $a(t)$ varies with
	$t$, and therefore the vacuum state of the theory is changing
	continuously. This effect leads to gravitational production
	of inflaton particles \cite{Birrell:1982ix}. Therefore we must
	expect a non-zero one-point function to be generated radiatively,
    reflecting the emergence of $\phi$ quanta from the vacuum.
    This issue is discussed in more detail in \S\ref{sec:one-point}
    below.

    The two-point expectation value defines the \emph{power spectrum}
    $P(k)$,
    \begin{equation}
	   \langle \delta\phi(\vect{k}_1) \delta\phi(\vect{k}_2) \rangle_\ast
	   = (2\pi)^3 \delta(\vect{k}_1 + \vect{k}_2) P_\ast(k_1) .
	\end{equation}
	The subscript `$\ast$' denotes evaluation at the time when the
	$\vect{k}$-mode under consideration left the horizon,
	which is unambiguous by virtue of momentum conservation which
	requires $k_1 = k_2$.
	At tree-level, $P_\ast(k) = H_\ast^2/2k^3$. It is often useful to
	work instead with the so-called dimensionless power spectrum,
	which is related to $P(k)$ by the rule $\Ps(k) = k^3 P(k)/2\pi^2$.
	
	The three-point expectation value defines the \emph{bispectrum},
	$B(k_1,k_2,k_3)$, and the four-point expectation value
	defines the \emph{trispectrum}, $T(\vect{k}_1,\vect{k}_2,
	\vect{k}_3,\vect{k}_4)$,
	\begin{equation}
		\langle \delta\phi(\vect{k}_1) \delta\phi(\vect{k}_2)
		        \delta\phi(\vect{k}_3)
		\rangle =
		(2\pi)^3 \delta( \sum_i \vect{k}_i )
		  B(k_1,k_2,k_3) ,
	\end{equation}
	and
	\begin{equation}
		\langle \delta\phi(\vect{k}_1) \delta\phi(\vect{k}_2)
		        \delta\phi(\vect{k}_3) \delta\phi(\vect{k}_4)
		\rangle =
		(2\pi)^3 \delta( \sum_i \vect{k}_i )
		  T(\vect{k}_1,\vect{k}_2,\vect{k}_3,\vect{k}_4) .
	\end{equation}
	Typically, $B$ and $T$ are proportional at tree-level to
	$\Ps^2$ and $\Ps^3$ respectively, multiplied by
	a momentum-dependent form-factor \cite{Maldacena:2002vr,
	Rigopoulos:2004gr,Rigopoulos:2004ba,Rigopoulos:2005xx,Rigopoulos:2005ae,
	Rigopoulos:2005us,Seery:2005gb,Seery:2006vu,Seery:2006js}.
	
	\section{Quantum corrections}
	\label{sec:quantum}
	
	\subsection{Loop corrections from Schwinger integrals}
	\label{sec:quantum-oneloop}
	
	\paragraph{Schwinger's formalism.}
	The appropriate formalism for computing expectation values in
	any quantum field theory was outlined by Schwinger
	\cite{Schwinger:1960qe}. Consider the vacuum expectation value of
	any observable $O$, observed at some time $t_\ast$,
	and computed in some theory with light scalar fields $\{ \phi^\alpha \}$.
	By inserting a complete set of states at
	any time $t_{\fin} > t_\ast$, this expectation value can be written
	\begin{equation}
		\langle \Omega | O | \Omega \rangle_\ast =
		\int [\d \varphi^\alpha] \langle \Omega |
		\phi_{\fin}^\alpha = \varphi^\alpha \rangle
		\langle \phi_{\fin}^\alpha = \varphi^\alpha |
		O |
		\Omega \rangle_\ast ,
		\label{eq:expectation}
	\end{equation}
	where $| \Omega \rangle$ is the vacuum state at $t \rightarrow
	-\infty$,
	the subscript `$\ast$' indicates that the fields in the
	expectation value are evaluated at $t_\ast$, and
	$\phi_{\fin}$ denotes $\phi$ evaluated at $t_{\fin}$. The integral
	$\int [\d \varphi^\alpha]$ is taken over all three-dimensional field
	configurations
	at $t_{\fin}$.
	Each factor in the product of transition
	amplitudes on the right-hand side of \eref{eq:expectation}
	can be expressed using the conventional Feynman path integral formula
	\cite{Weinberg:1995mt,Rivers:1987hi,Kleinert:2004ev},
	$\langle \phi_{\fin} = \varphi | O | \Omega \rangle =
	\int [\d\phi^\alpha]_{\Omega}^{\varphi} \; O \,
	\exp{ \im S }$, where $S$ is the action functional and the
	integral is taken over all field configurations which begin in
	the state $| \Omega \rangle$ and end in the state
	$| \phi_{\fin} = \varphi \rangle$. These boundary conditions on
	$\phi$ are schematically denoted by
	the limits $\Omega$ and $\varphi$ attached to $[\d \phi]$.
	
	\paragraph{The interacting vacuum.}
	\label{sec:interacting-vacuum}
	In order to evaluate
	such integrals by the usual Feynman diagram expansion it is necessary
	to remove these boundary conditions, so that we
	integrate over all $\phi$ unrestrictedly.
	We follow the analysis of Weinberg \cite{Weinberg:2005vy}.
	To remove the restriction
	that the field must begin in the vacuum state one can integrate over
	all $\phi$ obeying an \emph{arbitrary} boundary condition at $t \rightarrow
	-\infty$, after multiplying the integrand by the
	vacuum wavefunctional,
	$\Psi[\psi] = \langle \phi(-\infty) = \psi | \Omega
	\rangle$. This has the desired effect of restricting the integral to field
	configurations which begin in the correct vacuum. The exact
	expression for $\Psi$ depends on what we assume about $| \Omega \rangle$,
	but because the theory is supposed to be free as $t \rightarrow -\infty$
	it must be a gaussian in the fields
	\cite{Weinberg:1995mt,Weinberg:2005vy}. Therefore we assume
	\begin{equation}
		\fl
		\Psi[\psi] \propto \prod_\alpha
		\exp\Big\{ - \frac{1}{2} \int \frac{\d^3 q \, \d^3 r}{(2\pi)^3} \;
		\delta(\vect{q} + \vect{r}) \Omega_\alpha(q)
		\psi^\alpha(\vect{q}) \psi^\alpha(\vect{r})
		\Big\} \equiv
		\prod_\alpha
		\exp\Big\{ - \frac{1}{2}
		  (\psi^\alpha,\Omega_\alpha \psi^\alpha) \Big\} ,
	\end{equation}
	for some set of weight functionals $\{ \Omega_\alpha (q) \}$,
	where $(\psi,\Omega \psi)$ is a convenient abbreviation for the integral.
	The expectation value \eref{eq:expectation} can therefore be written
	\cite{Weinberg:2005vy}
	\begin{eqnarray}
		\fl\nonumber
		\langle \Omega | O | \Omega \rangle \propto
		\Bigg( \prod_\alpha \int [\d \varphi^\alpha] \Bigg)
		\Bigg\{ \Bigg( \prod_\beta \int [\d \phi_-^\beta]^{\varphi} \Bigg)
			\exp \left( \im S[\phi_-] \right)
			\prod_\beta \exp \Big[ -\frac{1}{2}
				(\psi_-^\beta,\Omega_\beta \psi_-^\beta) \Big]
		\Bigg\}^{\dag} \\
		\hspace{0.75cm}
		\Bigg\{ \Bigg( \prod_\gamma \int [\d \phi_+^\gamma]^{\varphi} \Bigg)
			\; O \;
			\exp \left( \im S[\phi_+] \right)
			\prod_\gamma \exp \Big[ - \frac{1}{2}
		    	(\psi_+^\gamma,\Omega_\gamma \psi_+^\gamma) \Big]
		\Bigg\}
		\label{eq:expectation-decomposed}
	\end{eqnarray}
	where `$\dag$' denotes Hermitian conjugation,
	and the fields in the two path integrals have been differentiated
	by the addition of subscripts `$+$' and `$-$'.
	The overall constant of proportionality is irrelevant.
	Since \eref{eq:expectation-decomposed} requires an integral over final
	field configurations, it is possible to drop the restriction on the
	fields $\phi_\pm$ at $t_{\fin}$, provided we guarantee that the $+$
	and $-$ fields for each species
	share a common value at this time. This can be accommodated by
	inserting a $\delta$-function into the integrand which constrains
	the fields to agree \cite{Weinberg:2005vy}
	\begin{equation}
		\prod_\alpha
		\delta\Big\{ \phi_+^\alpha(t_{\fin}) - \phi_-^\alpha(t_{\fin}) \Big\}
		\propto
		\lim_{\varepsilon \rightarrow 0}
		\exp \Big\{ - \frac{1}{\varepsilon} \sum_\alpha \Big[
		  \phi_+^\alpha(t_{\fin}) - \phi_-^\alpha(t_{\fin}) \Big]^2 \Big\} ,
	\end{equation}
	where $\varepsilon$ is positive.
	Also, the action is real by assumption so
	the only effect of Hermitian conjugation
	in \eref{eq:expectation-decomposed} is to flip the sign of the
	$\im S$ term.
	
	\paragraph{Solution for propagators.}
	Suppose that the action corresponds to a free field,
	so that it can be written $S = (2\pi)^{-3} \int
	\d^3 k_1 \, \d^3 k_2 \, \d t_1 \, \d t_2 \,
	\phi(t_1,\vect{k}_1) \laplacian \phi(t_2,\vect{k}_2)/2$ for
	some differential kernel $\laplacian$.
	Eq.~\eref{eq:expectation-decomposed} can be written as an
	unrestricted path integral over the fields $\{ \phi_+, \phi_- \}$
	of the form
	\begin{equation}
		\fl
		\Bigg( \prod_\alpha \int [ \d \phi_+^\alpha \, \d \phi_-^\alpha ]
		\Bigg)
		\exp\Bigg\{
			\frac{\im}{2} \int \frac{\d^3 k_1 \, \d^3 k_2}{(2\pi)^3}
			\, \d t_1 \, \d t_2
			\sum_\alpha
			\Big[ \!\! \begin{array}{c}
				\phi_+^\alpha(t_1,\vect{k}_1) \\
				\phi_-^\alpha(t_2,\vect{k}_2)
			\end{array}\!\!\Big]^{\transpose}
			K_{12}^\alpha
			\Big[ \!\! \begin{array}{c}
				\phi_+^\alpha(t_1,\vect{k}_1) \\
				\phi_-^\alpha(t_2,\vect{k}_2)
			\end{array}\!\!\Big]
		\Bigg\} ,
		\label{eq:quadratic}
	\end{equation}
	where we have assumed that there are no linear couplings among
	the various species,
	$\transpose$ denotes a transpose,
	and $K_{12}^\alpha$ is the $(2\times 2)$ kernel
	\begin{equation}
		\fl
		K_{12}^\alpha \equiv \delta(\vect{k}_1+\vect{k}_2)
		\Bigg( \!\! \begin{array}{cc}
		  \laplacian_\alpha
		  + \frac{2\im}{\varepsilon} \dfin{1} \dfin{2}
		  + \im \delta_{1\infty}\delta_{2\infty}\Omega_\alpha &
		  - \frac{2\im}{\epsilon} \dfin{1} \dfin{2} \\
		  - \frac{2\im}{\epsilon} \dfin{1} \dfin{2} &
		  - \laplacian^\alpha
		  + \frac{2\im}{\varepsilon} \dfin{1} \dfin{2}
		  + \im \delta_{1\infty}\delta_{2\infty}\Omega_\alpha
		\end{array} \!\! \Bigg) .
		\label{eq:kernel}
	\end{equation}
	In Eq.~\eref{eq:kernel}, $\dfin{j}$ is the $\delta$-function
	$\delta(t_j - t_{\fin})$ and
	$\delta_{j\infty}$ is the $\delta$-function
	$\delta(t_j + \infty)$. We will also occasionally use the notation
	$\delta_{ij} \equiv \delta(t_i - t_j)$.
	The field propagator matrix for any particular species,
	consisting of propagators
	$\{ G_{++}, G_{+-}, G_{-+}, G_{--} \}$ which connect the $+$ and $-$
	fields,
	is found by inverting the quadratic term given in~\eref{eq:quadratic},
	\begin{equation}
		\fl
		\int \d t_2 \, \d^3 k_2 \, K_{12}
		\Bigg( \begin{array}{cc}
		  G_{++} & G_{+-} \\ G_{-+} & G_{--}
		\end{array} \Bigg)_{23} = \im (2\pi)^3
		\delta_{13}\delta(\vect{k}_1-\vect{k}_3)
		\Bigg( \begin{array}{cc}
		  1 & 0 \\ 0 & -1 \end{array} \Bigg) .
		\label{eq:inverse}
	\end{equation}
	The subscript `23' indicates that $G_{++}$ is a function
	of times and momenta in the form
	$G_{++}(t_2,\vect{k}_2;t_3,\vect{k}_3)$, etc.; and similarly for the
	other $G$.
	
	Eq.~\eref{eq:inverse} splits into coupled equations for
	$G_{++}$, $G_{--}$, $G_{+-}$ and $G_{-+}$. It will shortly become
	apparent that the doublets $(G_{++},G_{+-})$ and
	$(G_{--},G_{-+})$ are to be regarded
	as forming complex conjugate pairs, so half of these equations
	are related to the other half by complex conjugation.
	In the application of interest, $\laplacian$ is given to leading order
	in the slow-roll approximation by the laplacian of exact de Sitter space,
	\begin{equation}
		\laplacian_{12} =
		\frac{\partial}{\partial t_1}
		\frac{\partial}{\partial t_2}
		( a^3 \delta_{12} )
		+ (\vect{k}_1\cdot\vect{k}_2) a \delta_{12} .
	\end{equation}
	Write $G_{++}^{12} = (2\pi)^3 \delta(\vect{k}_1 + \vect{k}_2)
	\tilde{G}_{++}^{12}$.
	The $\tilde{G}_{++}$ equation reads
	\begin{equation}
		\fl
		\frac{\partial^2}{\partial t_1^2} \tilde{G}_{++}^{12} +
		3H(t_1) \frac{\partial}{\partial t_1} \tilde{G}_{++}^{12} +
		\frac{k_1^2}{a(t_1)^2} \tilde{G}_{++}^{12}
		- \frac{\im}{a^3} \delta_{1\infty} \Omega(k_1)
		\tilde{G}_{++}^{\infty 2} = - \frac{\im}{a^3} \delta_{12}
		\label{eq:plusplus} .
	\end{equation}
	$\tilde{G}_{+-}$ obeys the homogeneous version of~\eref{eq:plusplus},
	whereas $\tilde{G}_{--}$ obeys the complex conjugate of
	\eref{eq:plusplus} and $\tilde{G}_{-+}$ its homgeneous complex
	conjugate. In addition, Eq.~\eref{eq:inverse} gives
	$\tilde{G}_{+-}$ and $\tilde{G}_{-}$ the boundary conditions
	\begin{equation}
		\dfin{1} \tilde{G}_{+-}^{\fin 2} =
		\dfin{1} \tilde{G}_{--}^{\fin 2}
		\quad \mbox{and} \quad
		\dfin{1} \tilde{G}_{-+}^{\fin 2} =
		\dfin{1} \tilde{G}_{++}^{\fin 2} .
		\label{eq:green-boundary}
	\end{equation}
	
	\paragraph{Homogeneous equation.}
	Consider any solution, say $\tilde{G}$, to the homogeneous version
	of~\eref{eq:plusplus}. Any such solution is a function
	of the single variable $t_1$,
	which after changing to conformal time $\eta$
	can be written in the form $\tilde{G}(\eta) \equiv \zeta_k(\eta)/a(\eta)$
	for some function $\zeta(\eta)$ to be determined.
	[The dependence of the mixed propagators on a second time
	argument, $t_2$, enters only through the boundary
	conditions~\eref{eq:green-boundary}.] The mode function
	$\zeta_k$ must obey
	\begin{equation}
		\zeta_k'' + \Big\{
			k^2(1-2 \im \iota) - (aH)^2(2-\epsilon) \Big\} \zeta_k = 0
		\label{eq:mode}
	\end{equation}
	where $k$ is the common magnitude of $\vect{k}_1$ and $\vect{k}_2$,
	a prime $'$ denotes a derivative with respect to $\eta$,
	and $\iota$ satisfies
	\begin{equation}
		\iota \equiv \delta_{\eta\infty}
		\frac{\Omega(k)}{2 (a k)^2}
		\geq 0 .
	\end{equation}
	Eq.~\eref{eq:mode} is equivalent to the condition
	$\zeta_k'' + \{ k^2 - (aH)^2(2-\epsilon) \} \zeta_k = 0$
	almost everywhere,
	together with the boundary condition $(\zeta_k/a^2) \rightarrow 0$
	as $\eta \rightarrow -\infty$. Heuristically,
	this boundary condition can be
	accommodated most naturally by redefining the range of $\eta$ to include
	some evolution in imaginary time, $\eta \mapsto \eta(1 + \im \iota)$.
	Although strictly speaking this contour is singular, owing to the
	presence of the $\delta$-function, it can be approached as a limit
	of regular contours. It will be seen below that when integrations over
	$\eta$ are required the integrands in question are holomorphic.
	Therefore,
	any one of these regular contours suffices
	for calculation and we may as well take $\eta \mapsto
	\eta(1+\im \iota)$ for \emph{fixed} $\iota$.
	This prescription was used in
	Refs.~\cite{Maldacena:2002vr,Seery:2005gb,Seery:2006vu}
	to compute tree-level correlation functions of
	the $\{ \delta\phi^\alpha \}$ in the interacting vacuum.
	
	\paragraph{Propagator matrix.}
	One can now construct an explicit solution for $G_{++}$,
	which satisfies (in conformal time with arguments
	$\eta_1$ and $\eta_2$)
	\begin{equation}
		G_{++}^{12}(\vect{k}_1,\vect{k}_2) = (2\pi)^3
		\delta(\vect{k}_1+\vect{k}_2)
		\times
		\Bigg\{
		    \begin{array}{l@{\hspace{5mm}}l}
			    \xi_k(\eta_1,\eta_2) & \mbox{if $\eta_1 < \eta_2$} \\
			    \xi^\ast_k(\eta_1,\eta_2) & \mbox{if $\eta_2 < \eta_1$}
			\end{array}
		,
		\label{eq:green-plusplus}
	\end{equation}
	where `$\ast$' denotes complex conjugation, $k$ is the common
	magnitude of $\vect{k}_1$ and $\vect{k}_2$, and
	$\xi_k(\eta_1,\eta_2)$ is defined by
	\begin{equation}
		\xi_k(\eta_1,\eta_2) \equiv
		\im \frac{[W(\zeta^\ast_k,\zeta_k)]^{-1}}{a(\eta_1)a(\eta_2)}
		\zeta_k(\eta_1)\zeta^\ast_k(\eta_2)
		\label{eq:xi} ,
	\end{equation}
	in which $W(f,g)$ is
	the Wronskian $W(f,g) \equiv fg' - gf'$.
	Note that $\im [W(\zeta^\ast_k,\zeta_k)]^{-1}$
	is real and time-independent,
	in virtue of Abel's identity, but may depend on $k$.
	The propagator $G_{--}$ is obtained by complex conjugation
	of Eq.~\eref{eq:green-plusplus}; the mixed propagator
	$G_{+-}$ is obtained from a homogeneous equation and therefore
	is smooth at $\eta_1 = \eta_2$. The boundary condition
	\eref{eq:green-boundary} implies that it must satisfy
	\begin{equation}
		G_{+-}^{12}(\vect{k}_1,\vect{k}_2) = (2\pi)^3
		\delta(\vect{k}_1+\vect{k}_2) \xi_k(\eta_1,\eta_2)
		\label{eq:green-plusminus}
	\end{equation}
	and $G_{-+}$ is given by its complex conjugate.
	[Note that there is no ambiguity in deciding which
	propagator should be assigned to a mixed pair
	$\langle \delta\phi_+ \delta\phi_- \rangle$, because the
	mode $\zeta$ is always assigned to the argument of the
	$+$ field, and $\zeta^\ast$ is always assigned to the
	argument of the $-$ field.]
	
	The above analysis was carried out for a single field, but
	where more than one species of light field is present
	similar results apply, with a mode function $\zeta_k^\alpha$
	for each species which obeys a vacuum boundary condition
	of the form $(\zeta_k^\alpha/a^2) \rightarrow 0$ in the far past.
	The propagators which connect two
	fields of different species $\alpha$ and $\beta$ then obey
	analogues of~\eref{eq:green-plusplus} and~\eref{eq:green-plusminus}
	with the function $\xi_k$ replaced by a matrix $\xi^{\alpha\beta}_k$.
	If the fields do not couple linearly, then it follows that
	$\xi^{\alpha\beta}_k = \delta^{\alpha\beta}\xi_k$.
	
	\paragraph{One-vertex, one-loop amplitudes.}
	In the remainder of this paper, we shall be concerned with computing
	expectation values in which a set of $n$ external fields,
	$\phi(\vect{k}_n)$, observed at some time $\eta_\ast$
	and carrying momenta $\{ \vect{k}_n \}$,
	are paired with a single $(n+2)$-valent internal vertex
	with coupling constant $g$. Applying Schwinger's formula shows that the
	term in such an expectation value of leading order in $g$ is given by
	\begin{equation}
		\im (2\pi)^3 \int \frac{\d^3 q_1 \cdots \d^3 q_n \, \d^3 q_{n+1} \,
		\d^3 q_{n+2}}
		{(2\pi)^{3(n+2)}} \delta(\sum_{i = 1}^{n+2} \vect{q}_i)
		\int_{-\infty}^{\eta_{\fin}} \d \eta \; g
		M ,
		\label{eq:one-vertex}
	\end{equation}
	where $M$ is defined by
	\begin{eqnarray}
		\fl\nonumber
		M \equiv \left\langle		
		\phi_+^\alpha(\vect{k}_1) \cdots \phi_+^\beta(\vect{k}_n)
	    \phi_+^\gamma(\vect{q}_1) \cdots \phi_+^\delta(\vect{q}_{n})
		\phi_+^\rho(\vect{q}_{n+1}) \phi_+^\sigma(\vect{q}_{n+2})
	    \right\rangle \\ \hspace{-1cm}
	    \mbox{} -
	    \left\langle
	    \phi_+^\alpha(\vect{k}_1) \cdots \phi_+^\beta(\vect{k}_n)
	    \phi_-^\gamma(\vect{q}_1) \cdots \phi_-^\delta(\vect{q}_{n})
		\phi_-^\rho(\vect{q}_{n+1}) \phi_-^\sigma(\vect{q}_{n+2})
	    \right\rangle .
	    \label{eq:M-def}
	\end{eqnarray}
	Greek indices label the species of fields, which are here
	allowed to run over field derivatives as well as the fields themselves;
	the issue of obtaining a correct path integral for theories with
	derivative interactions causes no difficulties for the purposes of
	Eqs.~\eref{eq:one-vertex}--\eref{eq:M-def}, but will be taken up again
	in more detail in the next section.
	Any amplitude of the type given in~\eref{eq:one-vertex}--\eref{eq:M-def}
	is automatically of one-loop order, because the
	two field operators left over after all $n$
	external fields have been paired
	with $n$ of the vertex fields must contract amongst themselves,
	leaving a single unconstrained integral over momentum.
	
	The time integral in~\eref{eq:one-vertex} has been carried to some
	arbitrary late
	time $\eta_{\fin}$ which satisfies $\eta_{\fin} > \eta_\ast$. Using
	Eqs.~\eref{eq:green-plusplus}--\eref{eq:green-plusminus}
	together with their complex conjugates
	in Eq.~\eref{eq:one-vertex}, it follows that
	the expectation value can be written
	\begin{eqnarray}
		\fl\nonumber
		\im (2\pi)^3 \delta(\vect{k}_1 + \cdots + \vect{k}_n)
		\int \frac{\d^3 q}{(2\pi)^3} \int_{-\infty}^{\eta_\ast} \d \eta \;
		\xi_{k_1}^{\alpha\gamma}(\eta,\eta_\ast)
		\cdots
		\xi_{k_n}^{\beta\delta}(\eta,\eta_\ast)
		\xi_{q}^{\rho\sigma}(\eta,\eta) \\ \nonumber
		\mbox{} -
		\im (2\pi)^3 \delta(\vect{k}_1 + \cdots + \vect{k}_n)
		\int \frac{\d^3 q}{(2\pi)^3} \int_{-\infty}^{\eta_\ast} \d \eta \;
		\xi_{k_1}^{\alpha\gamma \ast}(\eta,\eta_\ast)
		\cdots
		\xi_{k_n}^{\beta\delta \ast}(\eta,\eta_\ast)
		\xi_{q}^{\rho\sigma \ast}(\eta,\eta) \\
		\mbox{} + \mbox{permutations} ,
		\label{eq:expectation-value}
	\end{eqnarray}
	in which the second term is the complex conjugate of the first,
	and all permutations likewise assemble into complex conjugate
	pairs. Observe that the internal term $\xi_{q}(\eta,\eta)$
	in the first line comes from pairing two $+$ fields, whereas in
	the second line it comes from pairing two $-$ fields.
	Eq.~\eref{eq:xi} shows that for the $\delta\phi$ propagator,
	$\xi_q(\eta,\eta)$ is real, so that $\xi_q(\eta,\eta)$
	and $\xi_q^\ast(\eta,\eta)$ are in fact equal.
	
	The part of the integration over times between $\eta_\ast$ and
	$\eta_\fin$ has cancelled out, since in this region
	Eq.~\eref{eq:green-plusminus} is given by the same expression
	as Eq.~\eref{eq:green-plusplus}, whereas in the region
	$\eta < \eta_\ast$ it is given by its complex conjugate.
	Note that for interactions which contain more than a single
	vertex and a single loop the process of deriving expressions
	such as~\eref{eq:expectation-value} using the path integral
	technology described above becomes increasingly
	cumbersome. For such interactions, some form of the diagrammatic
	operator formalism recently elaborated by Musso
	\cite{Musso:2006pt} is likely to prove superior
	(see also Ref. \cite{Meulen:2007ah}).
	
	\subsection{Theories with derivative interactions}
	\label{sec:quantum-derivatives}
	
	An important feature of the interactions~\eref{eq:cubic}
	and~\eref{eq:quartic} is that they include time derivatives
	of the perturbation, $\delta\dot{\phi}$
	\cite{Unruh:1998ic}. This means that the
	lagrangian can not be written in the canonical form
	$L(\delta\phi,\delta\dot{\phi}) = \frac{1}{2}
	\delta\dot{\phi}\laplacian\delta\dot{\phi} + V(\delta\phi)$
	(where the operator $\laplacian$ is field independent),
	in which there is only a quadratic dependence on $\delta\dot{\phi}$.
	As a result the textbook construction of the path integral
	formula based on $L$ does not work.
	
	In the standard construction one identifies a momentum, $\pi$,
	canonically conjugate to $\delta\phi$ and writes the lagrangian
	as a Legendre transformation of the hamiltonian function $H$,
	\begin{equation}
		L(\delta\phi,\delta\dot{\phi}) = \pi \delta\dot{\phi} -
		H(\delta\phi,\pi) .
	\end{equation}
	In the quantum theory $\delta\phi$ and $\pi$ cannot be
	specified simultaneously. Since it is $H$ that generates
	time evolution, when one constructs the path integral
	one naturally arrives at a functional integration that
	involves independent integrals over $\delta\phi$ and $\pi$.
	If $L$ depends at most quadratically on $\delta\dot{\phi}$
	then $H$ depends at most quadratically on $\pi$ and the
	momentum integral can be performed immediately. This has
	the effect of setting the value of $\pi$ equal to the one
	stipulated by Hamilton's equations and results in the
	standard lagrangian path integral formula \cite{Weinberg:1995mt}.
	However, when $H$ has a more complicated dependence on $\pi$
	the momentum integral must be treated more carefully.
	
	The properties of lagrangians with derivative interactions have
	been studied extensively in the context of the non-linear
	$\sigma$-model.
	(See, eg., Coleman~\cite{Coleman};
	a path integral treatment is given in Ref.~\cite{Weinberg:1995mt},
	whereas the canonical approach was followed in
	Ref.~\cite{Gerstein:1971fm}.) After inspection of
	Eqs.~\eref{eq:cubic}--\eref{eq:quartic} it is clear that no
	term contains as many as four time derivatives, although there
	are terms containing one, two or three. Let us parametrize
	a general action for a field $\theta$ with arbitrary interactions
	containing as many as three time derivatives in form
	\begin{equation}
		\fl
		S = (2\pi)^3 \int \d\eta \;
		\left( \frac{1}{2}\gamma_{\alpha\beta}
		       \dot{\theta}^\alpha\dot{\theta}^\beta -
		       \frac{1}{2}\hat{\delta}_{\alpha\beta}
		       \partial\theta^\alpha\partial\theta^\beta
		       - V(\theta) + \lambda_\alpha \dot{\theta}^\alpha
		       + \frac{1}{3} \omega_{\alpha\beta\gamma}
		       \dot{\theta}^\alpha \dot{\theta}^\beta
		       \dot{\theta}^\gamma
		\right) .
		\label{eq:derivative-action}
	\end{equation}
	In order to keep this and subsequent expressions manageable,
	Eq.~\eref{eq:derivative-action} has been written in an
	abbreviated ``de Witt'' notation, where contraction over indicies
	implies not only a summation over species, but also an integration
	over momentum variables with measure $\d^3 k/(2\pi)^3$.
	With these conventions the object $\hat{\delta}_{\alpha\beta}
	= \delta(\vect{k}_\alpha + \vect{k}_\beta)$
	is a ``pseudo-metric''
	on $\vect{k}$-space which is numerically identical to
	its index-raised counterpart, $\hat{\delta}^{\alpha\beta}$.%
		\footnote{Although it is tempting to regard $\hat{\delta}_{\alpha\beta}$
		as an object for raising and lowering indices, it is not a true
		metric because with our
		conventions, the object obtained by mixing indices,
		$\hat{\delta}_{\alpha\beta}\hat{\delta}^{\beta\gamma}$, is
		\emph{not} the
		identity operator $(2\pi)^3 \delta(\vect{k}_\alpha - \vect{k}_\gamma)$,
		although it is proportional to it.}
	Note that we have taken
	any interactions involving exactly two factors of $\dot{\theta}$ to
	be included with the kinetic term in $\gamma_{\alpha\beta}$.
	Moreover,
	without loss of generality $\gamma_{\alpha\beta}$ and
	$\omega_{\alpha\beta\gamma}$ can be supposed to be symmetric
	under exchange of their indices. We assume that
	$\gamma_{\alpha\beta}$ is invertible, with inverse
	$\gamma^{\alpha\beta}$.
	
	The momentum conjugate to $\dot{\theta}^\alpha$ is $\pi_\alpha$,
	\begin{equation}
		\pi_\alpha \equiv \frac{\delta S}{\delta \dot{\theta}^\alpha} =
		(2\pi)^3 \left( \gamma_{\alpha\beta} \dot{\theta}^\beta
		+ \lambda_\alpha
		+ \omega_{\alpha\beta\gamma}\dot{\theta}^\beta
		\dot{\theta}^\gamma \right )
	\end{equation}
	where we have used the assumed symmetry under index exchange to
	simplify this expression. In order to apply this formalism to
	the cubic and quartic interactions~\eref{eq:cubic}
	and~\eref{eq:quartic} it is only necessary to compute to
	$\Or(\theta^4)$, where we formally assume that $\pi \sim \theta$ in
	order of magnitude. Since there are no three-derivative
	interactions in $S_3$, this implies that
	$\omega = \Or(\theta)$ and it will be sufficient to work
	to leading order in $\omega$.
	To this order, the hamiltonian can be written
	\begin{eqnarray}
		\fl\nonumber
		H = \frac{1}{2} \frac{1}{(2\pi)^3} \gamma^{\alpha\beta}
		\pi_\alpha \pi_\beta + \frac{1}{2} (2\pi)^3
		\hat{\delta}_{\alpha\beta} \partial \theta^\alpha
		\partial \theta^\beta + \frac{1}{2}
		(2\pi)^3 \gamma^{\alpha\beta} \lambda_\alpha
		\lambda_\beta +
		(2\pi)^3 V \\
		\mbox{} - \frac{1}{3} \frac{1}{(2\pi)^6}
		\omega_{\alpha\beta\gamma} \gamma^{\alpha\rho}
		\gamma^{\beta\sigma} \gamma^{\gamma\tau}
		\pi_\rho \pi_\sigma \pi_\tau -
		\gamma^{\alpha\beta} \pi_\alpha \pi_\lambda .
	\end{eqnarray}
	This hamiltonian can be used to construct a path integral for
	$\theta$, giving
	\begin{equation}
		\int [\d \theta^\alpha \, \d \pi_\beta]
		\exp \left\{ \im \int \d\eta
		\left( \pi_\alpha \dot{\theta}^\alpha - H \right)
		\right\} .
		\label{eq:pathintegral}
	\end{equation}
	The fields $\dot{\theta}^\alpha$ and $\pi_\beta$ are now variables of
	integration, and therefore independent, so we are free to redefine the
	momentum field by a shift,
	\begin{equation}
		\pi_\alpha \mapsto (2\pi)^3 (
		\pi_\alpha + \chi_\alpha )
	\end{equation}
	with $\chi_\alpha$ chosen to eliminate the term in
	Eq.~\eref{eq:pathintegral} which is linear in $\pi_\alpha$,
	\begin{equation}
		\chi_\alpha \equiv
		\gamma_{\alpha\beta} \dot{\theta}^\beta + \lambda_\alpha
		+ \omega_{\alpha\beta\gamma}
		\dot{\theta}^\beta \dot{\theta}^\gamma .
	\end{equation}
	This shift leaves the path integral measure $[\d \pi_\beta]$ invariant.
	Having done so, one may rearrange terms to find a simplified
	path integral expression
	\begin{equation}
		\int [\d \theta^\alpha \, \d \pi_\beta]
		\; \exp \Big\{ \im (S_\theta + \Sghost) \Big\} ,
	\end{equation}
	where $S_\theta$ is the original $\theta$
	action~\eref{eq:derivative-action} with all derivative interactions
	in their original form, and $\Sghost$ is an effective action for
	the ``ghost'' field $\pi$,
	\begin{equation}
		\fl
		\Sghost = (2\pi)^3 \int \left(
		- \frac{1}{2} \gamma^{\alpha\beta} \pi_\alpha \pi_\beta +
		\omega_{\alpha\beta\gamma}\gamma^{\beta\sigma}
		\gamma^{\gamma\tau}\dot{\theta}^\alpha
		\pi_\sigma \pi_\tau +
		\frac{1}{3} \omega_{\alpha\beta\gamma}
		\gamma^{\alpha\rho}\gamma^{\beta\sigma}
		\gamma^{\gamma\tau}
		\pi_\rho \pi_\sigma \pi_\tau \right)
		\label{eq:ghost-action-total}
	\end{equation}
	The quanta associated with $\pi$ do not appear in physical
	states, although they couple to $\theta$ and so affect its
	expectation values when loop corrections are
	taken into account. This explains why it has been permissible
	to ignore such ghosts in previous tree-level calculations;
	the $\pi$ integral makes no contribution to tree-level
	expectation values. Note that unlike the more familiar
	Fadeev--Popov ghost, the $\pi$ field is a spacetime
	scalar and not a spin-1/2 fermion.
	
	Eq.~\eref{eq:ghost-action-total} is not yet in a form suitable
	for perturbative calculations. In particular the ghost
	kinetic term involves the inverse $\gamma^{\alpha\beta}$.
	This will be a complicated object even for relatively
	simple choices of $\gamma_{\alpha\beta}$, but it is
	pointless to compute beyond $\Or(\theta^4)$ since the
	action to which we wish to apply this formalism
	[namely Eqs.~\eref{eq:kinetic}--\eref{eq:quartic}]
	was truncated at this level. For a canonically normalised
	scalar field in an almost-de Sitter spacetime
	$\gamma_{\alpha\beta}$ can be written
	\begin{equation}
		\gamma_{\alpha\beta} =
		a^2 \delta(\vect{k}_\alpha + \vect{k}_\beta) +
		2 \Gamma_1(\vect{k}_\alpha,\vect{k}_\beta) +
		2 \Gamma_2(\vect{k}_\alpha,\vect{k}_\beta) + \cdots
		\label{eq:gamma}
	\end{equation}
	where the $\Gamma_m$ are taken to be $\ord(\theta^m)$,
	the factors of two have been inserted for future
	convenience, and `$\cdots$' denotes higher order terms
	which have been omitted.
	
	The identity operator with our conventions is
	$\delta^\gamma_\alpha \equiv
	(2\pi)^3\delta(\vect{k}_\alpha-\vect{k}_\beta)$.
	Therefore, the inverse $\gamma^{\alpha\beta}$ can be written
	\begin{equation}
		\gamma^{\alpha\beta} = (2\pi)^6 \left(
		\frac{1}{a^2}\delta(\vect{k}_\alpha+\vect{k}_\beta)
		+ \psi_1(\vect{k}_\alpha,\vect{k}_\beta)
		+ \psi_2(\vect{k}_\alpha,\vect{k}_\beta)
		\right) ,
		\label{eq:gamma-inverse}
	\end{equation}
	where the $\psi_m$ are taken to be $\ord(\theta^m)$.
	One can verify that the normalization in
	Eq.~\eref{eq:gamma-inverse} is correct, since when~\eref{eq:gamma}
	and~\eref{eq:gamma-inverse} are contracted together the
	$\ord(\theta^0)$ term becomes
	\begin{equation}
		\fl
		\int \frac{\d^3 k_\beta}{(2\pi)^3} \;
		a^2 \delta(\vect{k}_\alpha + \vect{k}_\beta)
		\cdot
		(2\pi)^6 \frac{1}{a^2}
		\delta(\vect{k}_\beta + \vect{k}_\gamma)
		=
		(2\pi)^3 \delta(\vect{k}_\alpha - \vect{k}_\gamma) =
		\delta^\gamma_\alpha .
	\end{equation}
	The $\ord(\theta)$ equation implies that $\psi_1$ satisfies
	\begin{equation}
		\psi_1(\vect{k}_\alpha,\vect{k}_\beta) \equiv -
		\frac{2}{a^4} \Gamma_1(-\vect{k}_\alpha,-\vect{k}_\beta) ,
		\label{eq:psione-def}
	\end{equation}
	whereas the $\ord(\theta^2)$ equation implies that $\psi_2$ satisfies
	\begin{equation}
		\fl
		\psi_2(\vect{k}_\alpha,\vect{k}_\beta) \equiv -
		\frac{2}{a^4} \Gamma_2(-\vect{k}_\alpha,-\vect{k}_\beta)
		- \frac{4}{a^4} \int \d^3 q \;
		\Gamma_1(-\vect{k}_\alpha,\vect{q})\Gamma_1(-\vect{q},-\vect{k}_\beta)
		.
		\label{eq:psitwo-def}
	\end{equation}
	The ghost action can therefore be written
	\begin{eqnarray}
		\fl\nonumber
		\Sghost = (2\pi)^9 \int \d\eta \; \Bigg\{
		- \frac{1}{2a^2} \hat{\delta}^{\alpha\beta}
		\pi_\alpha \pi_\beta - \psi_1^{\alpha\beta}
		\pi_\alpha \pi_\beta - \psi_2^{\alpha\beta}
		\pi_\alpha \pi_\beta
		+ \frac{(2\pi)^6}{a^4} \omega_{\alpha\beta\gamma}
		\hat{\delta}^{\beta\sigma} \hat{\delta}^{\gamma\tau}
		\dot{\theta}^\alpha \pi_\sigma \pi_\tau \\
		\hspace{1cm}
		\mbox{} +
		\frac{(2\pi)^{12}}{3a^6} \omega_{\alpha\beta\gamma}
		\hat{\delta}^{\alpha\rho}
		\hat{\delta}^{\beta\sigma}
		\hat{\delta}^{\gamma\tau}
		\pi_\rho \pi_\sigma \pi_\tau \Bigg\}
		\label{eq:ghost-action}
	\end{eqnarray}
	The first term is $\ord(\theta)^2$ and can be taken as the free-field
	part, whereas the remainder is $\Or(\theta^3)$
	and can be taken as the interaction term. In this form, the ghost
	action is suitable for perturbative evaluation.
	
	\subsection{Feynman rules for the interacting scalar/ghost theory}
	\label{sec:quantum-rules}
	
	We are now in a position to write down the Feynman rules for the
	$\delta\phi$ theory, including the effect of the ghost field.
	In this section we will not be obliged to carry out any of the
	complicated manipulations which characterized
	\S\ref{sec:quantum-derivatives}, and so we will revert to a notation
	in which momentum labels and integrals are written explicitly.
	
	The propagators for the pure $\delta\phi$ theory were written
	down in \S\ref{sec:quantum-oneloop}. The free part of the
	ghost action can be inverted immediately to find the ghost
	propagator. For the $+$ fields this gives
	\begin{equation}
		\langle \pi_+(\eta_1,\vect{k}_1) \pi_+(\eta_2,\vect{k}_2)
		\rangle = - \frac{\im}{(2\pi)^3} a(\eta_1)a(\eta_2)
		\delta(\eta_1 - \eta_2) \delta(\vect{k}_1 + \vect{k}_2) ,
		\label{eq:ghost-plusplus}
	\end{equation}
	from which the $--$ propagator can be obtained by complex conjugation.
	Eq.~\eref{eq:ghost-plusplus} is the propagator for a
	so-called \emph{static ultra-local} field
	\cite{Caianiello:1974pw,Rivers:1987hi}.
	Its $\vect{k}$ and $\eta$ dependence is constrained by the appearance
	of $\delta$-functions, so the ghost does not propagate:
	its purpose is to provide corrections to the \emph{vertices} of the
	theory which account for the presence of coincident time derivatives
	there.
	At one-loop order, we do not require the mixed propagator;
	the ghost field only appears in loops, but at one-loop the only
	mixed contractions involve pairings of external fields with
	internal fields and these cannot occur for the ghost. 
		
	In the one-loop, single-field vertex formula,
	Eq.~\eref{eq:expectation-value},
	the ghost propagator always appears in the role of the propagator
	evaluated at equal arguments,
	associated with the internal momentum $\vect{q}$.
	Note that the special simplification, which occurred for
	the $\delta\phi$ propagator, where
	$\xi_q(\eta,\eta)$ and $\xi_q^\ast(\eta,\eta)$ were equal,
	does not apply for the ghost field.
	
	It remains to identify the interaction terms $V$, $\lambda$,
	$\Gamma_{1}$, $\Gamma_2$ and $\omega$. Reading these off from
	Eqs.~\eref{eq:cubic}--\eref{eq:quartic} we obtain%
		\footnote{In \texttt{v1}--\texttt{2}
		of the \texttt{arXiv} version of this paper,
		and the version which subsequently
		appeared in JCAP, the interaction $\Gamma_2$ contained a sign
		error in its third term, which propagated through the remainder of
		the calculation.
		This caused
		the coefficient of $\ln k$ which appeared in
		Eq.~\eref{eq:total-loop} to be incorrect.
		I would like to thank P. Adshead,
		R. Easther and E. Lim for bringing this error to my
		attention.
		This error was corrected in Ref.~\cite{Adshead:2008gk}; however,
		owing to a typographical error in that reference,
		the final coefficient of $\ln k$ was again given incorrectly
		(P. Adshead, personal communication).}
	\begin{equation}
		\fl
		V = a^2 \frac{\dot{\phi}}{4H}
		\int \frac{\d^3 q_1 \, \d^3 q_2 \, \d^3 q_3}
		{(2\pi)^9} \; \delta(\sum_{i=1}^3 \vect{q}_i)
		\left\{ \prod_{j=1}^3 \delta\phi(\vect{q}_j) \right\}
		(\vect{q}_2 \cdot \vect{q}_3)
		\label{eq:V}
	\end{equation}
	\begin{equation}
		\fl
		\lambda(\vect{q}_1) = \frac{a}{4H}
		\int \frac{\d^3 q_2 \, \d^3 q_3 \, \d^3 q_4}{(2\pi)^9}
		\; \delta(\sum_{i=1}^4 \vect{q}_i)
		\left\{ \prod_{j=2}^4 \delta\phi(\vect{q}_j) \right\}
		(\vect{q}_2 \cdot \vect{q}_3)
		\frac{\sigma(\vect{q}_1,\vect{q}_4)}{q_{14}^2} ,
		\label{eq:lambda}
	\end{equation}
	\begin{equation}
		\fl
		\Gamma_1(\vect{q}_1,\vect{q}_2) = - a^2 \frac{\dot{\phi}}{4H}
		\int \frac{\d^3 q_3}{(2\pi)^3} \;
		\delta(\sum_{i=1}^3 \vect{q}_i)
		\delta\phi(\vect{q}_3)
		\left( 1 + 2 \frac{\sigma(\vect{q}_2,\vect{q}_3)}{q_1^2} \right)
		\label{eq:gammaone}
	\end{equation}
	\begin{eqnarray}
		\fl\nonumber
		\Gamma_2(\vect{q}_1,\vect{q}_2) = - a^2
		\int \frac{\d^3 q_3 \, \d^3 q_4}{(2\pi)^6} \;
		\delta(\sum_{i=1}^4 \vect{q}_i)
		\left\{ \prod_{j=3}^4 \delta\phi(\vect{q}_j) \right\} \\
		\mbox{} \times
		\left(
		\frac{\z(\vect{q}_1,\vect{q}_3)\cdot\z(\vect{q}_2,\vect{q}_4)}
		     {q_{13}^4 q_{24}^2}
		+ \frac{3}{4}
		\frac{\sigma(\vect{q}_1,\vect{q}_3)\sigma(\vect{q}_2,\vect{q}_4)}
		     {q_{13}^2 q_{24}^2}
		+ 2 \frac{\vect{q}_4 \cdot \z(\vect{q}_2,\vect{q}_3)}{q_{23}^4}
		\right)
		\label{eq:gammatwo}
	\end{eqnarray}
	\begin{equation}
		\fl
		\omega(\vect{q}_1,\vect{q}_2,\vect{q}_3) = - \frac{a}{4H}
		\int \frac{\d^3 q_4}{(2\pi)^3} \;
		\delta(\sum_{i=1}^4 \vect{q}_i)
		\delta\phi(\vect{q}_4)
		\frac{\sigma(\vect{q}_1,\vect{q}_4)}{q_{14}^2} ,
		\label{eq:omega}
	\end{equation}
	where $\vect{q}_{ij} = \vect{q}_i + \vect{q}_j$.
	The functions $\sigma$ and $\z$ are defined by
	\begin{equation}
		\sigma(\vect{a},\vect{b}) \equiv \vect{a}\cdot\vect{b} + b^2
	\end{equation}
	and
	\begin{equation}
		\z(\vect{a},\vect{b}) \equiv \sigma(\vect{a},\vect{b})\vect{a} -
		\sigma(\vect{b},\vect{a})\vect{b} .
	\end{equation}
	These are the momentum-space counterparts of
	Eqs.~\eref{eq:vectortwo}--\eref{eq:sigma}.
	Note that as written, Eqs.~\eref{eq:gammaone}--\eref{eq:gammatwo}
	for $\Gamma$ and Eq.~\eref{eq:omega} for $\omega$ are not
	symmetric under exchange of their arguments. For $\Gamma$ this is
	immaterial, because~\eref{eq:derivative-action}
	and~\eref{eq:ghost-action} show that it always appears in a
	symmetric contraction. On the other hand, $\omega$ does appear once
	in an asymmetric contraction, namely
	$\omega_{\alpha\beta\gamma}\dot{\theta}^\alpha
	\hat{\delta}^{\beta\sigma}\hat{\delta}^{\gamma\tau}
	\pi_\sigma \pi_\tau$.
	To avoid an unnecessarily tripling of the length
	of~\eref{eq:omega} we leave it in asymmetric form,
	carrying out an explicit symmetrization when computing
	amplitudes involving the asymmetric vertex.
	
	\paragraph{Diagrammatic representation.}
	Eqs.~\eref{eq:V}--\eref{eq:omega} lead to a rather complicated
	diagrammatic formalism in which the vertices produce a number of
	related terms, depending on the number of derivatives which apply
	to the lines entering the vertex. In order to keep track of these
	related contributions it is useful to introduce a refinement of
	the Feynman rules in which the lines of scalar propagators
	to which derivatives are applied are decorated with a dot.
	
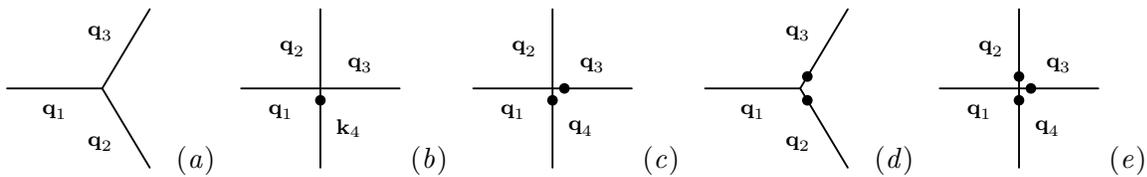
\begin{figure}
		\begin{center}
			\hfill
			\begin{fmfgraph*}(60,60)
				\fmfpen{0.8thin}
				\fmfleft{l}
				\fmfright{r1,r2}
				\fmf{plain,label=$\scriptsize\vect{q}_1$}{l,v}
				\fmf{plain,label=$\scriptsize\vect{q}_2$}{r1,v}
				\fmf{plain,label=$\scriptsize\vect{q}_3$}{r2,v}
			\end{fmfgraph*}
			\flabel{a}
			\hfill
			\begin{fmfgraph*}(60,60)
				\DerivativeInteractions
				\fmfpen{0.8thin}
				\fmfleft{l}
				\fmfright{r}
				\fmftop{t}
				\fmfbottom{b}
				\fmf{plain,label=$\scriptsize\vect{q}_1$}{l,v}
				\fmf{plain,label=$\scriptsize\vect{q}_2$}{t,v}
				\fmf{plain,label=$\scriptsize\vect{q}_3$}{r,v}
				\fmf{derivative,label=$\scriptsize\vect{k}_4$}{b,v}
			\end{fmfgraph*}
			\flabel{b}
			\hfill
			\begin{fmfgraph*}(60,60)
				\DerivativeInteractions
				\fmfpen{0.8thin}
				\fmfleft{l}
				\fmfright{r}
				\fmftop{t}
				\fmfbottom{b}
				\fmf{plain,label=$\scriptsize\vect{q}_1$}{l,v}
				\fmf{plain,label=$\scriptsize\vect{q}_2$}{t,v}
				\fmf{derivative,label=$\scriptsize\vect{q}_3$}{r,v}
				\fmf{derivative,label=$\scriptsize\vect{q}_4$}{b,v}
			\end{fmfgraph*}
			\flabel{c}
			\hfill
			\begin{fmfgraph*}(60,60)
				\DerivativeInteractions
				\fmfpen{0.8thin}
				\fmfleft{l}
				\fmfright{r1,r2}
				\fmf{plain,label=$\scriptsize\vect{q}_1$}{l,v}
				\fmf{derivative,label=$\scriptsize\vect{q}_2$}{r1,v}
				\fmf{derivative,label=$\scriptsize\vect{q}_3$}{r2,v}
			\end{fmfgraph*}
			\flabel{d}
			\hfill
			\begin{fmfgraph*}(60,60)
				\DerivativeInteractions
				\fmfpen{0.8thin}
				\fmfleft{l}
				\fmfright{r}
				\fmftop{t}
				\fmfbottom{b}
				\fmf{plain,label=$\scriptsize\vect{q}_1$}{l,v}
				\fmf{derivative,label=$\scriptsize\vect{q}_2$}{t,v}
				\fmf{derivative,label=$\scriptsize\vect{q}_3$}{r,v}
				\fmf{derivative,label=$\scriptsize\vect{q}_4$}{b,v}
			\end{fmfgraph*}
			\flabel{e}
			\hfill
			\mbox{}
		\end{center}
		\caption{Pure $\delta\phi$ vertices. A dot on a scalar line entering a vertex
		shows that a time derivative is applied to the field at the point of interaction.
		In terms of Eq.~\eref{eq:derivative-action}, the diagrams correspond to the
		vertices produced by ($a$) the potential $V$; ($b$) the $\lambda$ vertex;
		($c$) the $\Gamma_1$ vertex; ($d$) the $\Gamma_2$ vertex; and ($e$)
		the $\omega$ vertex.
		\label{fig:phi-diagrams}} 
	\end{figure}

\begin{figure}
		\begin{center}
			\hfill
			\begin{fmfgraph*}(60,60)
				\fmfpen{0.8thin}
				\fmfleft{l}
				\fmfright{r1,r2}
				\fmf{plain,label=$\scriptsize\vect{q}_1$}{l,v}
				\fmf{dashes,label=$\scriptsize\vect{q}_2$}{r1,v}
				\fmf{dashes,label=$\scriptsize\vect{q}_3$}{r2,v}
			\end{fmfgraph*}
			\flabel{a}
			\hfill
			\begin{fmfgraph*}(60,60)
				\fmfpen{0.8thin}
				\fmfleft{l}
				\fmfright{r}
				\fmftop{t}
				\fmfbottom{b}
				\fmf{plain,label=$\scriptsize\vect{q}_1$}{l,v}
				\fmf{plain,label=$\scriptsize\vect{q}_2$}{t,v}
				\fmf{dashes,label=$\scriptsize\vect{q}_3$}{r,v}
				\fmf{dashes,label=$\scriptsize\vect{q}_4$}{b,v}
			\end{fmfgraph*}
			\flabel{b}
			\hfill
			\begin{fmfgraph*}(60,60)
				\DerivativeInteractions
				\fmfpen{0.8thin}
				\fmfleft{l}
				\fmfright{r}
				\fmftop{t}
				\fmfbottom{b}
				\fmf{plain,label=$\scriptsize\vect{q}_1$}{l,v}
				\fmf{derivative,label=$\scriptsize\vect{q}_2$}{t,v}
				\fmf{dashes,label=$\scriptsize\vect{q}_3$}{r,v}
				\fmf{dashes,label=$\scriptsize\vect{q}_4$}{b,v}
			\end{fmfgraph*}
			\flabel{c}
			\hfill
			\begin{fmfgraph*}(60,60)
				\fmfpen{0.8thin}
				\fmfleft{l}
				\fmfright{r}
				\fmftop{t}
				\fmfbottom{b}
				\fmf{plain,label=$\scriptsize\vect{q}_1$}{l,v}
				\fmf{dashes,label=$\scriptsize\vect{q}_2$}{t,v}
				\fmf{dashes,label=$\scriptsize\vect{q}_3$}{r,v}
				\fmf{dashes,label=$\scriptsize\vect{q}_4$}{b,v}
			\end{fmfgraph*}
			\flabel{d}
			\hfill
			\mbox{}
		\end{center}
		\caption{Scalar/ghost vertices. Solid lines represent the scalar field
		$\delta\phi$, whereas dashed lines represent the ghost.
		A dot on a scalar line entering a vertex shows that a time derivative
		is applied to the field at the point of interaction. Time
		derivatives are never applied to ghost fields. In terms of
		Eq.~\eref{eq:ghost-action} the diagrams correspond
		to the vertices produced by ($a$) the $\psi_1 \pi^2$ interaction;
		($b$) the $\psi_2 \pi^2$ interaction;
		($c$) the $\omega \dot{\theta} \pi \pi$ interaction;
		and ($d$) the $\omega \pi^3$ interaction.
		\label{fig:ghost-diagrams}}
	\end{figure}
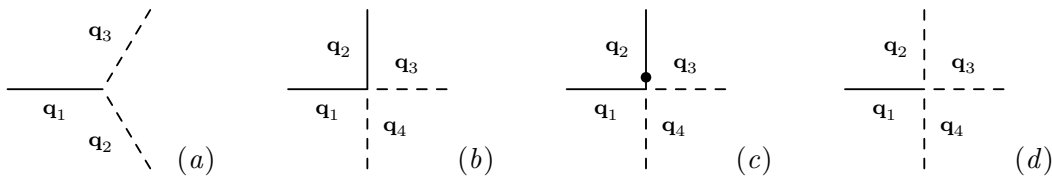

	For the pure $\delta\phi$ vertices, the resulting diagrams are
	depicted in Fig.~\ref{fig:phi-diagrams}. For the mixed
	$\delta\phi$/ghost vertices, the resulting diagrams are
	shown in Fig.~\ref{fig:ghost-diagrams}.
	
	\section{The one-point function}
	\label{sec:one-point}
	
	In \S\ref{sec:expectation-values}, we observed that at tree-level
	the one point function of $\delta\phi$ is zero,
	$\langle \delta\phi(\vect{k}) \rangle = 0$. This is not merely a
	question of convention; if the one-point function was \emph{not} zero
	then so-called `tadpole'
	diagrams such as Fig.~\eref{fig:one-point} would mean that
	$\delta\phi$ quanta would emerge from the vacuum.
	Conservation of momentum forces such particles to
	condense
	in the zero-momentum mode, and the accumulation
	of such particles causes the classical background field
	to change.
	Such an instability implies that any perturbation theory based
	on the original unstable vacuum state would not give meaningful
	answers. This problem can be avoided by ensuring that the vacuum
	which we take as the basis of our perturbation theory is stable,
	at least at tree-level.
	
	\begin{figure}
		\begin{center}
			\begin{fmfgraph*}(60,60)
				\fmfpen{0.8thin}
				\fmfleft{l}
				\fmfright{r}
				\fmf{plain,label=$\scriptsize\vect{q}=0$}{l,r}
				\fmfblob{0.15w}{r}
			\end{fmfgraph*}
		\end{center}
		\caption{Instability of the vacuum due to condensation.
		$\delta\phi$ particles emerge from the vacuum
		(represented by the hatched condensate) in a zero momentum state.
		The accumulation of such particles changes the homogeneous
		classical field configuration associated with the vacuum.
		\label{fig:one-point}}
	\end{figure}
	
	In an inflationary universe, the emergence of $\delta\phi$ quanta
	from the vacuum is exploited to produce small density fluctuations
	on superhorizon scales. Therefore we may expect to encounter
	some symptoms of vacuum instability when quantum corrections are
	taken into account. These symptoms manifest themselves as a
	radiatively generated one-point function,
	\begin{equation}
		\langle \delta\phi(\vect{k}) \rangle =
		(2\pi)^3 \delta(\vect{k}) O ,
	\end{equation}
	where $O \neq 0$ is a dimensionless quantity.
	Although it is not the observable in which we are principally
	interested, the present section is devoted to a calculation of $O$.
	This is important for two reasons. The first is that it provides
	a consistency check on $\delta N$ calculations
	\cite{Seery:2005gb,Seery:2006vu,Lyth:2005fi,Boubekeur:2005fj,
	Seery:2006js,Byrnes:2006vq,Byrnes:2007tm} which typically assume
	$O = 0$, even beyond tree level. The second is that it allows us
	to develop some aspects of the calculational formalism in a
	simpler setting than the computation of the two-point function.
	
	\subsection{Ghost diagrams}
	Consider the one-point function associated with some wavenumber
	$\vect{k}$. We aim to compute this at the time $\eta_\ast$
	when $\vect{k}$
	crosses the horizon, which is roughly defined by the condition
	$-k \eta_\ast = 1$. Eventually conservation of momentum will force
	us to set $\vect{k} = 0$, but in order to regularize the calculation
	we compute for finite $k$ and then study the limit $k \rightarrow 0$.

	We deal first with the diagrams which contain a ghost loop.
	There is only one such diagram, which arises from the
	$\psi_1\pi^2$ coupling,
	\begin{equation-diagram}
		\langle \delta\phi(\vect{k}) \rangle_\ast \hspace{1mm}
		\subseteq \hspace{2mm}
		\parbox[t]{15mm}{
			\begin{fmfgraph}(40,6)
				\fmfpen{0.6thin}
				\fmfleft{l}
				\fmfright{r}
				\fmf{plain}{l,v1}
				\fmf{dashes,left}{v1,v2}
				\fmf{dashes,left}{v2,v1}
				\fmf{phantom,tension=50}{v2,r}
			\end{fmfgraph}
		} .
	\end{equation-diagram}
	This diagram makes a contribution to $\langle
	\delta\phi(\vect{k}) \rangle_\ast$ equal to
	\begin{eqnarray}
		\fl\nonumber
		\langle \delta\phi(\vect{k}) \rangle_\ast \supseteq
		- (2\pi)^3 \delta(\vect{k}) \int_{-\infty}^{\eta_\ast}\d\eta
		\int \frac{\d^3 q}{(2\pi)^3} \frac{H_\ast H}{2k^3}
		\delta(0)(1-\im k \eta)
		\e{\im k \eta} \frac{\dot{\phi}}{4H}
		\left(1 + 2 \frac{\sigma(-\vect{q},\vect{k})}{q^2} \right)
		\\
		\mbox{} + \mbox{complex conjugate} ,
	\end{eqnarray}
	where the symbol `$\supseteq$' indicates that
	$\langle \delta\phi(\vect{k}) \rangle$ contains the
	indicated contribution (among others), and
	in deference to the vacuum prescription outlined in
	\S\ref{sec:quantum-oneloop} we should deform the contour of
	the $\eta$ integral to include some evolution in imaginary
	time for large $|\eta|$. In this region the exponential
	factor is strongly decaying (cf. the discussion in
	\S\ref{sec:interacting-vacuum}),
	so there is very little contribution to the integral from
	very early times; if $\eta_\ast$ is not too late,
	the integral receives its dominant contributions from
	times around horizon crossing, where $\eta \sim -1/k$.
	We may therefore approximate the slowly varying factors
	$H_\ast H$ and $\dot{\phi}/H$ by their values
	at the time of horizon crossing, which are equal to
	$H_\ast^2$ and $\dot{\phi}_\ast/H_\ast$ respectively.
	In this simple example the $\eta$ integral and the
	integral over the internal momentum $\vect{q}$ factorize,
	leaving a final result
	\begin{eqnarray}
		\fl\nonumber
		\langle \delta\phi(\vect{k}) \rangle_\ast \supseteq
		-(2\pi)^3 \delta(\vect{k}) P_\ast(k)
		\frac{\dot{\phi}_\ast}{2H_\ast}
		\int_{-\infty}^{\eta_\ast} \d \eta \;
		\delta(0)(1-\im k\eta)\e{\im k\eta}
		\int \frac{\d^3 q}{(2\pi)^3} \;
		\left( 1 + 2 \frac{\sigma(\vect{q},-\vect{k})}{q^2}
		\right)
		\\ \mbox{} + \mbox{complex conjugate}
		\label{eq:onepoint-ghost}
	\end{eqnarray}
	where $P_\ast(k)$ is the tree-level power spectrum
	evaluated at $\eta_\ast$.
	The object $\delta(0)$ is the $\eta$ delta-function
	evaluated at zero argument, and is badly divergent.
	In the present case, however, this is not material.
	The $\eta$ integral can be rotated to imaginary time,
	leaving a result which is purely imaginary. Hence,
	although divergent, this diagram makes no contribution
	to the one-point function.
	
	\subsection{Pure $\delta\phi$ diagrams}
	Now consider the pure $\delta\phi$ diagrams.
	
	\paragraph{The vacuum prescription and renormalization.}
	In these diagrams, as above,
	early times make almost no contribution to the $\eta$
	integral, so that slowly varying quantities such as
	$H$ and $\dot{\phi}/H$ can be
	evaluated at $\eta_\ast$. A generic pure-$\delta\phi$
	contribution to $O$ will then take the form
	\begin{equation}
		\fl
		O \supseteq \im P_\ast(k) \frac{\dot{\phi}_\ast}
		{4H_\ast} \int_{-\infty}^{\eta_\ast} \d \eta
		\int \frac{\d^3 q}{(2\pi)^3} \;
		\e{\im k \eta}
		\Sigma ,
		\label{eq:Sigma} +
		\mbox{complex conjugate} ,
	\end{equation}
	where $\Sigma$ is a $\vect{k}$- and $\vect{q}$-dependent
	quantity which is to be calculated.
	In evaluating $\Sigma$ we will encounter instances
	where a $\delta\phi$ propagator begins and ends at the
	same vertex, giving it coincident time arguments.
	We will choose to set such a propagator equal to
	\begin{equation}
		\langle \delta\phi(\vect{q}_1,\eta)
		        \delta\phi(\vect{q}_2,\eta)
		\rangle = (2\pi)^3 \delta(\vect{q}_1 + \vect{q}_2)
		\frac{H^2}{2q^3} (1 + q^2 \eta^2) ,
		\label{eq:loop-propagator}
	\end{equation}
	where the exponential factors $\e{\im q \eta}
	\e{-\im q \eta}$ have cancelled among themselves,
	and $q$ is the common magnitude of $\vect{q}_1$ and $\vect{q}_2$.
	
	The discussion of vacuum boundary conditions in
	\S\ref{sec:quantum-oneloop} emphasized that the fields
	which participate in the
	Schwinger path integral must be chosen to begin in the appropriate
	interacting vacuum, and that this could be achieved heuristically
	by deforming the contour of integration to include some evolution
	in imaginary time. In view of this, one may question
	whether~\eref{eq:loop-propagator} is the correct choice,
	or whether it should be modified to read%
	  \footnote{It makes no difference if we allow $\eta$ to develop a small
	  imaginary component in the prefactor $|1-\im q \eta|^2$,
	  since the exponential term is so strongly decaying for
	  large $|\eta|$.}
	\begin{equation}
		\fl
		\langle \delta\phi(\vect{q}_1,\eta)
		        \delta\phi(\vect{q}_2,\eta)
		\rangle \overset{?}{=}
		(2\pi)^3 \delta(\vect{q}_1 + \vect{q}_2)
		\frac{H^2}{2q^3} |1 - \im q \eta|^2
		\e{-2q \Im(\eta)}, \quad \mbox{where $\Im(\eta) > 0$} .
		\label{eq:loop-modified-propagator}
	\end{equation}
	This would apparently have the very desirable effect of decoupling our
	prediction for $\langle \delta\phi(\vect{k}) \rangle$ from
	any details of the deep
	ultraviolet regime where $q \rightarrow \infty$,
	because~\eref{eq:loop-modified-propagator} is strongly decaying
	in this limit owing to the exponential factor.
	Therefore one might have some reservations that the choice
	of~\eref{eq:loop-propagator} would introduce
	unphysical divergences, arising from an incorrect
	treatment of the vacuum. On the other hand,
	Eq.~\eref{eq:loop-modified-propagator}
	has the undesirable feature that it leads to a non-holomorphic
	integrand. This means that it would be necessary to rescind the
	possibility of contour rotation in evaluating the $\eta$
	integral. A loop amplitude computed
	using~\eref{eq:loop-modified-propagator} would therefore depend
	sensitively upon the supposedly arbitrary value we assign to
	$\Im(\eta)$, which in turn depends on the structure of the vacuum
	at past infinity.
	It is not clear how the resulting amplitude should be interpreted.
	
	This situation can be understood as follows. It was explained in
	\S\ref{sec:interacting-vacuum} that the trick of contour rotation can
	be expected to account reliably for the vacuum boundary
	conditions only when the integrand is holomorphic.
	This is automatically true
	at tree-level, as argued in \S\ref{sec:interacting-vacuum},
	and gives completely
	unambiguous results which are independent of the details of
	physics in the deep ultraviolet.
	At one-loop level the situation is different.
	If we adopt Eq.~\eref{eq:loop-propagator}, then the loop amplitude
	depends on the ultraviolet parameter $\Im(\eta)$.
	If we adopt
	the holomorphic expression Eq.~\eref{eq:loop-propagator} for the
	propagator, then the result is free of any dependence on $\Im(\eta)$
	but there is instead a prospect
	of sensitivity to the details of ultraviolet
	physics from the $\vect{q}$ integral.
	Although we can choose from which source this sensitivity arises,
	we cannot evade it altogether---as we should expect.
	However, any
	divergences associated with the limit $q \rightarrow
	\infty$ can be subtracted by conventional methods of
	renormalization, after which the contour-rotation prescription
	gives a finite, contour-independent result which correctly incorporates
	the vacuum boundary conditions. In what follows we adopt this
	prescription.
	
	This leaves open the question of how the $q$-integral should be regulated.
	Since the Einstein action is supposed to be an effective
	theory of gravity for energies less than the Planck
	scale $\Planck \approx 10^{18} \, \mbox{GeV}$, and
	inflation is usually supposed to occur at energies at
	least a few orders of magnitude \emph{less} than $\Planck$,
	one might imagine applying a cutoff on the loop momenta
	of order the Planck scale. However this in itself is
	ambiguous since the Planck scale, unlike the speed of light,
	is not a Lorentz invariant and varies between locally
	inertial frames. In particular, the comoving Planck scale
	at a given instant $\eta$ is given by $a(\eta)\Planck$.
	Therefore a momentum cutoff of this form entangles the
	$\eta$ and $\vect{q}$ integrations, and for this reason it seems
	preferable to use a method of regularization, such as
	dimensional regularization, which does not depend on the explicit
	use of a cutoff.
	In the present paper we
	will compute expectation values using a fixed momentum
	cutoff both in the infrared and ultraviolet.
	In the case of the one-point function it may be checked
	that when the ultraviolet region has been discarded
	both dimensional regularization and a fixed momentum cutoff yield
	comparable predictions for the leading
	infrared divergences.
	
	This is sufficient for the purposes of the present paper,
	since it is the behaviour in the
	infrared rather than the ultraviolet which is of principal
	interest in a cosmological context.
	Divergences in the ultraviolet come from the
	behaviour of the fields at high energies and small scales.
	Such small scale modes exist far inside the horizon, where
	the equivalence principle suggests that
	flat spacetime quantum field theory is expected to be a good
	approximation. The subtraction of these modes has recently been
	considered by Finelli {\etal}, who argue that no
	special treatment is required for the power spectrum
	\cite{Finelli:2007fr} (see also Ref.~\cite{Ashoorioon:2004wd}).
	On the other hand, the infrared behaviour
	comes from low energies and large scales, where the
	field modes are well outside the horizon. On such scales,
	flat spacetime field theory is a very poor approximation and
	we are obliged to take account of the gravitational background.
	
	This does not preclude the appearance of new ultraviolet divergences
	in our expectation values.
	Indeed, many of the integrals we shall encounter
	\emph{do} contain ultraviolet divergences, of which the
	ultraviolet divergent quantity $\delta(0)$ which appears
	in Eq.~\eref{eq:onepoint-ghost} is an example. Such divergences
	do not interfere with our ability to perform accelerator
	or laboratory particle physics
	experiments on earth, which are characterized by
	time- and length-scales that are small compared to the
	expansion time-scale and horizon length-scale of the universe.
	On such small scales the ultraviolet divergences
	we shall encounter (none of which are present in the $\delta\phi$
	theory in Minkowski space)
	are presumably subdominant with respect to divergences
	from the pure matter theory and therefore do not interfere with
	our ability to
	perform terrestrial experiments, or with the success of the
	principle of equivalence.
	
	\paragraph{Zero-derivative interactions.}
	We now return to the $\delta\phi$ diagrams. It is
	simplest to classify these diagrams
	according to the number of
	derivatives applied to propagators entering the vertex.
	
	There is a zero-derivative interaction from the vertex
	in Fig.~\eref{fig:phi-diagrams}\flabel{a},
	\begin{equation-diagram}
		\langle \delta\phi(\vect{k}) \rangle_\ast \hspace{1mm}
		\supseteq \hspace{2mm}
		\parbox[t]{15mm}{
			\begin{fmfgraph}(40,6)
				\fmfpen{0.6thin}
				\fmfleft{l}
				\fmfright{r}
				\fmf{plain}{l,v1}
				\fmf{plain,left}{v1,v2}
				\fmf{plain,left}{v2,v1}
				\fmf{phantom,tension=50}{v2,r}
			\end{fmfgraph}
		} .
	\end{equation-diagram}
	This term makes a contribution to $\Sigma$ which equals
	\begin{equation}
		\Sigma \supseteq - \left( \frac{1}{2q\eta^2}
		  - \frac{\vect{k}\cdot\vect{q}}{q^3 \eta^2} \right)
		  (1-\im k \eta) (1 + q^2 \eta^2).
	\end{equation}
	The term involving $\vect{k}\cdot\vect{q}$ is not rotationally
	invariant and disappears in the integral over $\vect{q}$.
	Let us introduce a fixed ultraviolet cutoff
	$\Lambda$ and infrared cutoff $\mu$.
	Evaluating the $\vect{q}$ and
	$\eta$ integrals as described above gives
	\begin{equation}
		O_\ast \supseteq - P_\ast \frac{\dot{\phi}_\ast}{4H_\ast}
		\frac{1}{4\pi^2 k} ( \Lambda^4 - \mu^4 ) .
		\label{eq:onepoint-one}
	\end{equation}
	
	\paragraph{Two derivatives, both derivatives on internal leg.}
	The next class of diagrams contain two derivative operators, and
	divide naturally
	into two sorts: those where the derivatives are applied to both
	ends of the internal loop, and those where one derivative is
	applied to the loop but the other is applied to the external leg.
	
	The first sort give rise to diagrams of the form
	\begin{equation-diagram}
		\langle \delta\phi(\vect{k}) \rangle_\ast \hspace{1mm}
		\supseteq \hspace{2mm}
		\parbox[t]{15mm}{
			\begin{fmfgraph}(40,6)
				\DerivativeInteractions
				\fmfpen{0.6thin}
				\fmfleft{l}
				\fmfright{r}
				\fmf{plain}{l,v1}
				\fmf{derivative,right}{v2,v1}
				\fmf{derivative,left}{v2,v1}
				\fmf{phantom,tension=50}{v2,r}
			\end{fmfgraph}
		} ;
	\end{equation-diagram}
	such diagrams contribute an amount to $\Sigma$ corresponding to
	\begin{equation}
		\Sigma \supseteq - \frac{q}{2} (1-\im k \eta)
		\left( 1 + 2 \frac{\sigma(\vect{q},-\vect{k})}{q^2} \right) .
	\end{equation}
	Evaluating the integrals by the method described above, one
	arrives at
	\begin{equation}
		O_\ast \supseteq - P_\ast(k) \frac{\dot{\phi}_\ast}{4H_\ast}
		\frac{1}{4\pi^2 k}
		( \Lambda^2 - \mu^2 ) ( 4 k^2 + \Lambda^2 + \mu^2 ) .
		\label{eq:onepoint-two}
	\end{equation}
	
	\paragraph{Two derivatives, single derivative on internal leg.}
	The final class of diagrams contain a single derivative
	on the internal line, and apply the remaining derivative to
	the external leg. These diagrams are of the form
	\begin{equation-diagram}
		\langle \delta\phi(\vect{k}) \rangle_\ast \hspace{1mm}
		\supseteq \hspace{2mm}
		\parbox[t]{15mm}{
			\begin{fmfgraph}(40,6)
				\DerivativeInteractions
				\fmfpen{0.6thin}
				\fmfleft{l}
				\fmfright{r}
				\fmf{derivative}{l,v1}
				\fmf{plain,right}{v2,v1}
				\fmf{small_derivative,left}{v2,v1}
				\fmf{phantom,tension=50}{v2,r}
			\end{fmfgraph}
		}
		\label{eq:onepoint-zero}
	\end{equation-diagram}
	and contribute to $\Sigma$ according to the rule
	\begin{equation}
		\Sigma \supseteq - \frac{k^2}{2q}
		(1-\im q \eta) \left( 2 + 2 \frac{\sigma(-\vect{k},\vect{q})}
		{q^2} \right)
	\end{equation}
	This class of diagrams makes a contribution to $O_\ast$ which
	equals
	\begin{equation} 
		O_\ast \supseteq -P_\ast(k) \frac{\dot{\phi}_\ast}{4H_\ast}
		\frac{1}{4\pi^2 k} \left(
			\frac{8k}{3} ( \Lambda^3 - \mu^3 )
			+ 4 k^2 ( \Lambda^2 - \mu^2 )
		\right) .
		\label{eq:onepoint-three}
	\end{equation}
	
	\subsection{Infrared behaviour}
	We now collect terms from
	Eqs.~\eref{eq:onepoint-one}, \eref{eq:onepoint-two}
	and~\eref{eq:onepoint-three}.
	The result shows only power law divergences,
	which could have been anticipated from the outset because the tadpole
	loop must be independent of $k$.
	Accordingly, the $\mu \rightarrow 0$ limit is perfectly regular,
	and yields
	\begin{equation}
		\langle \delta\phi(\vect{k}) \rangle_{\ast} =
		-(2\pi)^3 \delta(\vect{k})
		\frac{H_\ast \dot{\phi}_\ast}{16 \pi^2} \left(
			\frac{\Lambda^4}{k^4} +
			\frac{4}{3} \frac{\Lambda^3}{k^3}
			+ 4 \frac{\Lambda^2}{k^2}
		\right) .
		\label{eq:onepoint-final}
	\end{equation}
	This is purely divergent.
	Whatever renormalization scheme we choose, all these power-law
	divergences must be subtracted.
	The result may leave a $k$-independent remainder,
	but even if this is true the zero-momentum
	limit $k \rightarrow 0$ will be finite.
	
	In practice, inflation
	does not last for an indefinite number of e-folds and the region of the
	universe described by the inflationary patch will not be unboundedly large.
	One should identify the $\vect{k}=0$ mode with the spatial average of
	$\delta\phi$ within this patch. If $\langle \delta \phi(\vect{k})$
	is non-zero after renormalization,
	this spatial average can be absorbed into a redefinition
	of the background field $\phi(t)$ by enforcing the
	condition $\langle \delta\phi(\vect{k}) \rangle = 0$,
	as discussed (for example) in Refs.
	\cite{Boyanovsky:2005sh,Boyanovsky:2005px,Sloth:2006az,Sloth:2006nu}.
	It follows that when $\delta\phi$ is defined in this way
	one may take $O = 0$, as usually assumed.
	
	\section{The two-point function}
	\label{sec:two-point}
	
	We now turn to the central purpose of this paper, the
	computation of the leading loop correction to the
	two-point function
	$\langle \delta\phi(\vect{k}_1) \delta\phi(\vect{k}_2)
	\rangle_\ast$.
	
	It is first necessary to decide which classes of diagrams are
	to be included in the computation.
	In general, the one-loop correction to the two-point function
	of the $\delta\phi$ will be given by a sum of diagrams, the
	leading terms of which are of the form
	\begin{center}
		loop correction $\supseteq$ \hspace{1mm}
		\parbox{20mm}{\begin{fmfgraph*}(60,40)
			\fmfleft{l}
			\fmfright{r}
			\fmf{plain}{l,v}
			\fmf{plain}{r,v}
			\fmf{plain}{v,v}
		\end{fmfgraph*}}
		\hspace{1mm} + \hspace{1mm}
		\parbox{20mm}{\begin{fmfgraph*}(60,40)
			\fmfleft{l}
			\fmfright{r}
			\fmf{plain,tension=2}{l,v1}
			\fmf{plain,tension=2}{r,v2}
			\fmf{plain,left}{v1,v2}
			\fmf{plain,left}{v2,v1}
		\end{fmfgraph*}}
		\hspace{1mm}
		+ $\cdots$ .
	\end{center}
	Eqs.~\eref{eq:cubic}--\eref{eq:quartic} show that the leading
	contribution from the first diagram is $\Or(\epsilon^0)$
	in slow-roll, whereas the leading contribution from the second diagram is
	$\Or(\epsilon)$. Therefore, provided $\epsilon \ll 1$ and there are
	no large logarithms which can compensate for small slow-roll
	parameters, the expectation value will be dominated by the lowest-order
	part of the first diagram. This is \emph{opposite} to the case considered
	by Sloth \cite{Sloth:2006az,Sloth:2006nu}, where a large logarithm was
	used to compensate for the smallness of $\epsilon$. In this regime
	the first diagram will be dominated by its \emph{subleading}
	slow-roll part, and for consistency one should also take into account
	sub-leading slow-roll terms from the second diagram, and possibly
	from other sources.
	In the present paper, we wish to use the slow-roll approximation
	to simplify the calculation and therefore we will retain only the
	contribution from the leading part of the first diagram.
	The question of when this is a good approximation, together with
	a more general analysis of any possible large logarithms,
	will be postponed to another publication \cite{Seery:2007wf}.
	
	\subsection{Ghost diagrams}
	The relevant vertices here come from the $\psi_2 \pi\pi$
	and $\omega\delta\dot{\phi}\pi\pi$ terms in the ghost
	action. There is no contribution from the
	$\omega \pi\pi\pi$ interaction because this involves
	three ghost fields, which must appear in loops, and
	at one-loop order there is always one ghost field which
	is left unpaired. Therefore this term can be disregarded,
	although it would play a role in a two- or higher-loop
	calculation.
	To determine the $\psi_2$ contribution explicitly,
	consider
	Eq.~\eref{eq:psitwo-def} which gives $\psi_2$ in terms of
	the known functions $\Gamma_1$ and $\Gamma_2$. We are
	computing to leading order in slow-roll, so the term
	involving $\Gamma_1^2$ can be discarded, because
	Eq.~\eref{eq:gammaone} shows that it is proportional to the
	slow-roll parameter $\epsilon \sim \dot{\phi}^2/H^2$,
	whereas the leading terms in the fourth-order interaction
	are $\ord(\epsilon^0)$.
	
	The relevant ghost diagrams are
	\begin{equation-diagram}
		\langle \delta\phi(\vect{k}_1) \delta\phi(\vect{k}_2)
		\rangle_\ast \hspace{1mm} \supseteq \hspace{2mm}
		\parbox[t]{15mm}{
			\begin{fmfgraph}(40,6)
				\fmfpen{0.6thin}
				\fmfleft{l}
				\fmfright{r}
				\fmf{plain}{l,v}
				\fmf{dashes}{v,v}
				\fmf{plain}{r,v}
			\end{fmfgraph}
		}
		+
		\parbox[t]{15mm}{
			\begin{fmfgraph}(40,6)
				\DerivativeInteractions
				\fmfpen{0.6thin}
				\fmfleft{l}
				\fmfright{r}
				\fmf{small_derivative}{l,v}
				\fmf{dashes}{v,v}
				\fmf{plain}{r,v}
			\end{fmfgraph}
		} .
	\end{equation-diagram}
	Both these diagrams are purely imaginary and cancel between
	the $++$ and $--$ propagators in exactly the same manner
	described in \S\ref{sec:one-point} for the computation of the
	one-point function.
	
	The ghost diagrams have therefore entirely cancelled out
	in both the one- and two-point functions. This leads to expressions
	which agree with those reported in Refs.~\cite{Sloth:2006az,Sloth:2006nu}.
	However, one should not immediately conclude that the ghost diagrams
	always sum to zero. Although this issue deserves more detailed
	attention, at two-loop order and above
	one can presumably expect
	the factors of $\im$ to combine to give non-vanishing
	contributions. This will apparently
	occur whenever there are an even number of
	ghost propagators in the diagram.
	
	\subsection{Pure $\delta\phi$ diagrams}
	As in the one-point calculation, it is convenient to
	classify the pure $\delta\phi$ diagrams according to the number
	of derivatives they contain.
	
	\paragraph{Single derivative.}
	There are no zero-derivative interactions, because the
	gravitational interactions responsible for generating
	the vertices in Fig.~\ref{fig:phi-diagrams} make no
	contribution to the potential at $\ord(\delta\phi^4)$,
	and the cubic contribution which is generated would
	make a contribution to the loop correction which is
	subleading in slow-roll.
	
	The first non-trivial term contains a single
	derivative, which can be applied to an internal or external
	line,
	\begin{equation-diagram}
		\langle \delta\phi(\vect{k}_1) \delta\phi(\vect{k}_2)
		\rangle_\ast \hspace{1mm} \supseteq \hspace{2mm}
		\parbox[t]{15mm}{
			\begin{fmfgraph}(40,6)
				\DerivativeInteractions
				\fmfpen{0.6thin}
				\fmfleft{l}
				\fmfright{r}
				\fmf{small_derivative}{l,v}
				\fmf{plain}{v,v}
				\fmf{plain}{r,v}
			\end{fmfgraph}
		}
		+
		\parbox[t]{15mm}{
			\begin{fmfgraph}(40,6)
				\DerivativeInteractions
				\fmfpen{0.6thin}
				\fmfleft{l}
				\fmfright{r}
				\fmf{plain}{l,v}
				\fmf{derivative}{v,v}
				\fmf{plain}{r,v}
			\end{fmfgraph}
		} .
	\end{equation-diagram}
	As in the case of the one-point function, it is useful to parametrize
	the contribution each diagram makes to $\langle \delta\phi(\vect{k}_1)
	\delta\phi(\vect{k}_2) \rangle_\ast$ in terms of a function $\Pi$,
	which is defined by
	\begin{equation}
		\fl
		\langle \delta\phi(\vect{k}_1) \delta\phi(\vect{k}_2) \rangle_\ast
		= \im (2\pi)^3 P_\ast(k)^2 \int \d \eta \int \frac{\d^3 q}{(2\pi)^3}
		\; \e{2\im k \eta} \Pi + \mbox{complex conjugate},
	\end{equation}
	where $P_\ast(k)^2$ is the square of the tree-level power spectrum, and
	the quantity $\Pi$ (to be calculated in this section)
	depends on the external momenta $\{ \vect{k}_1,
	\vect{k}_2 \}$ and the loop momentum $\vect{q}$.
	
	The class of diagrams where the derivative is applied to the
	external leg makes a contribution to $\Pi$ which corresponds to
	\begin{eqnarray-diagram}
		\fl\nonumber
		\parbox[t]{15mm}{
			\begin{fmfgraph}(40,6)
				\DerivativeInteractions
				\fmfpen{0.6thin}
				\fmfleft{l}
				\fmfright{r}
				\fmf{small_derivative}{l,v}
				\fmf{plain}{v,v}
				\fmf{plain}{r,v}
			\end{fmfgraph}
		}
		: \hspace{3mm}
		\Pi \supseteq
		(1-\im k \eta)(1+q^2\eta^2)\left( \frac{k^2}{4q^3}
			(\vect{q}\cdot\vect{k}_2)
			\frac{\sigma(-\vect{k}_1,\vect{q})}{|\vect{k}_1 + \vect{q}|^2}
			+ \frac{k^2}{8q} \frac{\sigma(-\vect{k}_1,\vect{k}_2)}{k_{12}^2}
		\right) \\
		\mbox{} + [\vect{k}_1 \leftrightharpoons \vect{k}_2] ,
		\label{eq:twopoint-single-one}
	\end{eqnarray-diagram}
	
	\noindent
	where $[\vect{k}_1 \leftrightharpoons \vect{k}_2]$ denotes the same term
	with $\vect{k}_1$ and $\vect{k}_2$ interchanged. The ratio
	$\sigma(-\vect{k}_1,\vect{k}_2)/k_{12}^2$ is obviously singular when
	the momentum conservation condition $\vect{k}_1 + \vect{k}_2 = 0$ is
	enforced and must be treated carefully to avoid an unphysical
	divergence. Consider the non-singular quotient $\sigma(\vect{a},
	\vect{b})/|\vect{a}+\vect{b}|^2$ where $\vect{a}$ and $\vect{b}$
	approach $\vect{k}$ and $-\vect{k}$ respectively,
	\begin{equation}
		\lim_{\epsilon, \delta \rightarrow 0}
		\frac{\sigma(\vect{k} + \vect{\delta},-\vect{k}+\vect{\epsilon})}
		     {|\vect{\epsilon} + \vect{\delta}|^2} =
		\lim_{\epsilon, \delta \rightarrow 0}
		\frac{\vect{k}\cdot(\vect{\delta} + \vect{\epsilon}) +
		      \epsilon^2 + \vect{\delta}\cdot\vect{\epsilon}}
		     {\delta^2 + \epsilon^2 + 2\vect{\delta}\cdot
		      \vect{\epsilon}}
		\label{eq:sigma-regulated}
	\end{equation}
	This is not symmetric between $\vect{\delta}$ and
	$\vect{\epsilon}$, because $\sigma$ is not a symmetric function
	of its arguments; as a result, the limits do not commute.
	Moreover,
	as $\delta$ and $\epsilon$ approach zero the numerator
	of~\eref{eq:sigma-regulated} vanishes linearly, as fast as
	$\Or(\epsilon,\delta)$, whereas the denominator
	is vanishing quadratically, like $\Or(\epsilon^2,\delta^2)$.
	Therefore~\eref{eq:twopoint-single-one} is na\"{\i}vely
	divergent. In fact, the value of~\eref{eq:sigma-regulated}
	depends on what is assumed about $\vect{k}\cdot\vect{\delta}$
	and $\vect{k}\cdot\vect{\epsilon}$; if we demand that the
	limit is approached along a sequence of vectors of magnitude
	$k = |\vect{k}|$ then it follows that
	$|\vect{k} + \vect{\delta}| = |-\vect{k} + \vect{\epsilon}|
	= k$ and therefore
	\begin{equation}
		\vect{k}\cdot\vect{\epsilon} = \frac{\epsilon^2}{2}
		\quad \mbox{and} \quad
		\vect{k}\cdot\vect{\delta} = -\frac{\delta^2}{2} .
		\label{eq:symmetric-k-limit}
	\end{equation}
	With this choice, Eq.~\eref{eq:sigma-regulated} evaluates
	to $1/2$ and the limits become commuting. This prescription
	was used implicitly in \S4.2 of Ref.~\cite{Seery:2006vu},
	but there does not seem to be any compelling reason to
	demand that the limit is approached along such a specific
	sequence of vectors.
	Fortunately a catastrophic divergence is averted,
	since Eq.~\eref{eq:twopoint-single-one} requires symmetrization
	over the exchange $\vect{k}_1 \leftrightharpoons \vect{k}_2$.
	The problematic term $\vect{k}\cdot(\vect{\delta} +
	\vect{\epsilon})$ is antisymmetric under this exchange and
	cancels out of the expectation value~\eref{eq:twopoint-single-one},
	leaving a finite limit. The result of this procedure gives
	the same answer as if we had adopted
	Eq.~\eref{eq:symmetric-k-limit}, which can be regarded
	as a justification for the analysis
	presented in Ref.~\cite{Seery:2006vu}.
	
	After performing the symmetrization over
	$\vect{k}_1$ and $\vect{k}_2$ and
	integrating over $\vect{q}$ and $\eta$, this class
	of diagrams make a contribution to the two-point function
	of the form
	\begin{equation}
		\fl
		\langle \delta\phi(\vect{k}_1) \delta\phi(\vect{k}_2)
		\rangle_\ast \supseteq (2\pi)^3 \delta(\vect{k}_1 +
		\vect{k}_2) \frac{P_\ast(k)^2}{\pi^2}
		\left( - \frac{3}{16} k^3 \ln k - \frac{1}{120} k^3
		+ \cdots \right)
		\label{eq:twopoint-one}
	\end{equation}
	where `$\cdots$' denotes ultraviolet power law divergences
	which have
	been omitted.
	
	Now consider the diagrams in which the derivative is applied
	to the internal line. Such diagrams contribute to $\Pi$
	according to
	\begin{eqnarray-diagram}
		\fl\nonumber
		\parbox[t]{15mm}{
			\begin{fmfgraph}(40,6)
				\DerivativeInteractions
				\fmfpen{0.6thin}
				\fmfleft{l}
				\fmfright{r}
				\fmf{plain}{l,v}
				\fmf{derivative}{v,v}
				\fmf{plain}{r,v}
			\end{fmfgraph}
		}
		: \hspace{3mm}
		\Pi \supseteq
		(1-\im k \eta)^2(1-\im q \eta )\left( \frac{\vect{k}_1 \cdot \vect{q}}{4q}
			\frac{\sigma(-\vect{q},-\vect{k}_2)}{|\vect{k}_2 + \vect{q}|^2}
			+ \frac{k^2}{8q} \frac{\sigma(\vect{q},-\vect{q})}{|\vect{q}-\vect{q}|^2}
		\right) \\
		\mbox{} + [\vect{k}_1 \leftrightharpoons \vect{k}_2] ,
		\label{eq:twopoint-single-two}
	\end{eqnarray-diagram}
	
	\noindent
	This class of diagrams contains a similar ill-defined ratio,
	$\sigma(\vect{q},-\vect{q})/|\vect{q}-\vect{q}|$.
	Consider Eq.~\eref{eq:sigma-regulated} again, with $\vect{k}$
	replaced by $\vect{q}$. Although there is no longer any
	injunction to symmetrize over $\vect{q} \leftrightharpoons
	-\vect{q}$, the non-rotationally-invariant part
	$\vect{q}\cdot(\vect{\delta}+\vect{\epsilon})$ will vanish
	underneath the integral and does not give rise to any
	divergence. In order to assign a definite value to the remaining
	limit, we must assume something about $\vect{\delta}$
	and $\vect{\epsilon}$. Since $\vect{q}$ is merely a
	variable of integration and can be freely replaced by $-\vect{q}$,
	we assume that $\sigma(\vect{q},-\vect{q})$ is to be
	regularized by taking its symmetric part. With this prescription,
	the ratio $\sigma(\vect{q},-\vect{q})/|\vect{q}-\vect{q}|$
	evaluates to 1/2.
	
	Symmetrizing over $\vect{k}_1$ and $\vect{k}_2$, and
	omitting ultraviolet power laws, we find
	\begin{equation}
		\fl
		\langle \delta\phi(\vect{k}_1) \delta\phi(\vect{k}_2)
		\rangle_\ast \supseteq (2\pi)^3 \delta(\vect{k}_1 +
		\vect{k}_2) \frac{P_\ast(k)^2}{\pi^2}
		\left( - \frac{1}{8} k^3 \ln k + \frac{3}{20} k^3
		+ \cdots \right)
		\label{eq:twopoint-two}
	\end{equation}
	
	\paragraph{Two derivatives.}
	We have now exhausted all diagrams with only a single
	derivative. The next set of diagrams all involve two
	derivatives and break naturally into three sets:
	the first class includes all diagrams with the
	derivatives applied to both external legs of the
	two-point function; the second set includes all diagrams where
	one derivative is applied to an external leg while the other
	applies to the internal propagator; and the third set
	includes all diagrams with both derivatives applied to the internal
	propagator:
	\begin{equation-diagram}
		\langle \delta\phi(\vect{k}_1) \delta\phi(\vect{k}_2)
		\rangle_\ast \hspace{1mm} \supseteq \hspace{2mm}
		\parbox[t]{15mm}{
			\begin{fmfgraph}(40,6)
				\DerivativeInteractions
				\fmfpen{0.6thin}
				\fmfleft{l}
				\fmfright{r}
				\fmf{small_derivative}{l,v}
				\fmf{plain}{v,v}
				\fmf{small_derivative}{r,v}
			\end{fmfgraph}
		}
		+
		\parbox[t]{15mm}{
			\begin{fmfgraph}(40,6)
				\DerivativeInteractions
				\fmfpen{0.6thin}
				\fmfleft{l}
				\fmfright{r}
				\fmf{small_derivative}{l,v}
				\fmf{derivative}{v,v}
				\fmf{plain}{r,v}
			\end{fmfgraph}
		}
		+
		\parbox[t]{15mm}{
			\begin{fmfgraph}(40,6)
				\DerivativeInteractions
				\fmfpen{0.6thin}
				\fmfleft{l}
				\fmfright{r}
				\fmf{plain}{l,v}
				\fmf{double_derivative}{v,v}
				\fmf{plain}{r,v}
			\end{fmfgraph}
		}
	\end{equation-diagram}
	
	Consider first the set of diagrams with both derivatives on
	the external legs. We obtain
	\begin{equation-diagram}
		\fl
		\parbox[t]{15mm}{
			\begin{fmfgraph}(40,6)
				\DerivativeInteractions
				\fmfpen{0.6thin}
				\fmfleft{l}
				\fmfright{r}
				\fmf{small_derivative}{l,v}
				\fmf{plain}{v,v}
				\fmf{small_derivative}{r,v}
			\end{fmfgraph}
		}
		: \hspace{3mm}
		\Pi \supseteq
		- \frac{k^4}{2q^3} ( 1 + q^2 \eta^2 ) Q ,
	\end{equation-diagram}
	where $Q$ is the quantity
	\begin{eqnarray}
		\fl\nonumber
		Q \equiv
		\frac{\z(-\vect{k}_1,\vect{q})\cdot\z(-\vect{k}_2,\vect{q})}
		{|\vect{q}-\vect{k}_1|^4|\vect{q}+\vect{k}_2|^2} +
		\frac{3}{4}
		\frac{\sigma(-\vect{k}_1,\vect{q})\sigma(-\vect{k}_2,-\vect{q})}
		{|\vect{q}-\vect{k}_1|^2|\vect{q}+\vect{k}_2|^2} -
		2
		\frac{\vect{q}\cdot\z(-\vect{k}_2,\vect{q})}
		{|\vect{q}-\vect{k}_2|^4} \\
		\mbox{} + [\vect{k}_1 \leftrightharpoons \vect{k}_2] .
	\end{eqnarray}
	Unlike the previous examples,
	none of the ratios which appear in $Q$ are
	ill-defined. However, this result can still be
	simplified using the
	symmetry properties of $\z$ and $\sigma$. In particular,
	we observe that $\sigma$ is a quadratic form, and therefore
	\begin{equation}
		\sigma(-\vect{a},-\vect{b}) = \sigma(\vect{a},\vect{b})
		\quad \mbox{and} \quad
		\z(-\vect{a},-\vect{b}) = \z(\vect{b},\vect{a}) .
	\end{equation}
	These identities can be used together with the obvious
	antisymmetry of $\z$ [{\ie}, $\z(\vect{a},\vect{b}) = -
	\z(\vect{b},\vect{a})$].
	After performing the symmetrization
	over $\vect{k}_1$ and $\vect{k}_2$, $Q$ can be reduced to the
	simpler form
	\begin{equation}
		Q = - 2 \frac{\z(\vect{q},\vect{k})^2}{|\vect{q}+\vect{k}|^6} +
		\frac{3}{2} \frac{\sigma(\vect{k},\vect{q})^2}
		{|\vect{q}+\vect{k}|^4} -
		4\frac{\vect{q}\cdot\z(\vect{k},\vect{q})}{|\vect{q}+\vect{k}|^4} .
	\end{equation}
	In this expression, $\vect{k}$ can be taken to be either
	$\vect{k}_1$ or $\vect{k}_2$; after integration, the result depends
	only on the magnitude $k$ and not its orientation, and we obtain
	\begin{equation}
		\fl
		\langle \delta\phi(\vect{k}_1) \delta\phi(\vect{k}_2)
		\rangle_\ast \supseteq (2\pi)^3 \delta(\vect{k}_1 +
		\vect{k}_2) \frac{P_\ast(k)^2}{\pi^2}
		\left( \frac{13}{48} k^3 \ln k - \frac{4}{9} k^3
		+ \cdots \right)
		\label{eq:twopoint-three} .
	\end{equation}
	
	The set of diagrams with one derivative on an external leg
	and one derivative on the internal propagator are the most
	complicated. To evaluate them, we write
	\begin{equation-diagram}
		\fl
		\parbox[t]{15mm}{
			\begin{fmfgraph}(40,6)
				\DerivativeInteractions
				\fmfpen{0.6thin}
				\fmfleft{l}
				\fmfright{r}
				\fmf{small_derivative}{l,v}
				\fmf{derivative}{v,v}
				\fmf{plain}{r,v}
			\end{fmfgraph}
		}
		: \hspace{3mm}
		\Pi \supseteq
		- \frac{k^2}{2q} (1-\im k \eta)(1 - \im q \eta) R ,
	\end{equation-diagram}
	where $R$ can be expressed as
	\begin{equation}
		\fl\nonumber
		R \equiv
		4 \frac{\z(\vect{q},\vect{k})^2}{|\vect{q}+\vect{k}|^6}
		+ 3 \frac{\sigma(\vect{q},\vect{k})\sigma(\vect{k},\vect{q})}
		{|\vect{q}+\vect{k}|^4} + \frac{3}{2} -
		4 \frac{\vect{k}\cdot\z(\vect{k},\vect{q})}{|\vect{q}+\vect{k}|^4} -
		4 \frac{\vect{q}\cdot\z(\vect{q},\vect{k})}{|\vect{q}+\vect{k}|^4} .
	\end{equation}
	in which we have used the symmetry properties of $\z$ and $\sigma$,
	and the same convention that $\vect{k}$ may be chosen as either
	$\vect{k}_1$ or $\vect{k}_2$ applies. In arriving at this expression
	for $Q$, we have discarded a number of contributions of the form
	$\vect{X}\cdot\{ \z(-\vect{k},\vect{k}) + \z(\vect{k},-\vect{k}) \}$
	for some vector $\vect{X}$, which may itself require regularization.
	However, no matter how we choose to regularize the bracket $\{ \cdots \}$,
	the antisymmetry of $\z$ guarantees that it sums to zero, and therefore
	that such contributions cancel out of the observable expectation
	value.

	After integration, one obtains
	\begin{equation}
		\fl
		\langle \delta\phi(\vect{k}_1) \delta\phi(\vect{k}_2)
		\rangle_\ast \supseteq (2\pi)^3 \delta(\vect{k}_1 +
		\vect{k}_2) \frac{P_\ast(k)^2}{\pi^2}
		\left( - k^3 \ln k + \frac{5}{6} k^3
		+ \cdots \right)
		\label{eq:twopoint-four} .
	\end{equation}
	
	The final class of diagrams of this type involve both derivatives
	applied to the internal propagator,
	\begin{equation-diagram}
		\fl
		\parbox[t]{15mm}{
			\begin{fmfgraph}(40,6)
				\DerivativeInteractions
				\fmfpen{0.6thin}
				\fmfleft{l}
				\fmfright{r}
				\fmf{plain}{l,v}
				\fmf{double_derivative}{v,v}
				\fmf{plain}{r,v}
			\end{fmfgraph}
		}
		: \hspace{3mm}
		\Pi \supseteq
		- \frac{q}{2} (1-\im k \eta)^2
		\left( - 2 \frac{\z(\vect{q},\vect{k})^2}{|\vect{q}+\vect{k}|^6} +
		\frac{3}{2} \frac{\sigma(\vect{q},\vect{k})^2}{|\vect{q}+\vect{k}^4} -
		4 \frac{\vect{k}\cdot\z(\vect{q},\vect{k})}{|\vect{q}+\vect{k}|^4}
		\right),
	\end{equation-diagram}
	which does not require regularization. After integration, one obtains
	\begin{equation}
		\fl
		\langle \delta\phi(\vect{k}_1) \delta\phi(\vect{k}_2)
		\rangle_\ast \supseteq (2\pi)^3 \delta(\vect{k}_1 +
		\vect{k}_2) \frac{P_\ast(k)^2}{\pi^2}
		\left( \frac{19}{48} k^3 \ln k - \frac{23}{180} k^3
		+ \cdots \right)
		\label{eq:twopoint-five} .
	\end{equation}

	\paragraph{Three derivatives.}
	The only remaining class of diagrams are those containing three
	derivatives at the vertex. These diagrams break into two groups:
	those in which one end of the internal propagator is free of a derivative,
	and those in which an external leg is free of a derivative:
	\begin{equation-diagram}
		\langle \delta\phi(\vect{k}_1) \delta\phi(\vect{k}_2)
		\rangle_\ast \hspace{1mm} \supseteq \hspace{2mm}
		\parbox[t]{15mm}{
			\begin{fmfgraph}(40,6)
				\DerivativeInteractions
				\fmfpen{0.6thin}
				\fmfleft{l}
				\fmfright{r}
				\fmf{small_derivative}{l,v}
				\fmf{derivative}{v,v}
				\fmf{small_derivative}{r,v}
			\end{fmfgraph}
		}
		+
		\parbox[t]{15mm}{
			\begin{fmfgraph}(40,6)
				\DerivativeInteractions
				\fmfpen{0.6thin}
				\fmfleft{l}
				\fmfright{r}
				\fmf{small_derivative}{l,v}
				\fmf{double_derivative}{v,v}
				\fmf{plain}{r,v}
			\end{fmfgraph}
		}
	\end{equation-diagram}
	Both types give rise to comparatively simple expressions.
	For the first we obtain
	\begin{equation-diagram}
		\fl
		\parbox[t]{15mm}{
			\begin{fmfgraph}(40,6)
				\DerivativeInteractions
				\fmfpen{0.6thin}
				\fmfleft{l}
				\fmfright{r}
				\fmf{small_derivative}{l,v}
				\fmf{derivative}{v,v}
				\fmf{small_derivative}{r,v}
			\end{fmfgraph}
		}
		: \hspace{3mm}
		\Pi \supseteq
		\frac{k^4}{8q} \eta^2 (1-\im q \eta)
		\left( 4 \frac{\sigma(\vect{k},\vect{q})}{|\vect{q}+\vect{k}|^2}
			+ 1
		\right) ;
	\end{equation-diagram}
	after integration this class of diagrams give contributions totalling
	\begin{equation}
		\fl
		\langle \delta\phi(\vect{k}_1) \delta\phi(\vect{k}_2)
		\rangle_\ast \supseteq (2\pi)^3 \delta(\vect{k}_1 +
		\vect{k}_2) \frac{P_\ast(k)^2}{\pi^2}
		\left( -\frac{1}{24} k^3 \ln k + \frac{1}{18} k^3
		+ \cdots \right)
		\label{eq:twopoint-six} .
	\end{equation}
	On the other hand, for the second type of diagram we obtain
	\begin{equation-diagram}
		\fl
		\parbox[t]{15mm}{
			\begin{fmfgraph}(40,6)
				\DerivativeInteractions
				\fmfpen{0.6thin}
				\fmfleft{l}
				\fmfright{r}
				\fmf{small_derivative}{l,v}
				\fmf{double_derivative}{v,v}
				\fmf{plain}{r,v}
			\end{fmfgraph}
		}
		: \hspace{3mm}
		\Pi \supseteq
		\frac{k^2 q}{8} \eta^2 (1-\im k \eta)
		\left( 4 \frac{\sigma(\vect{q},\vect{k})}{|\vect{q}+\vect{k}|^2}
			+ 1
		\right) .
	\end{equation-diagram}
	After integration, we find
	\begin{equation}
		\fl
		\langle \delta\phi(\vect{k}_1) \delta\phi(\vect{k}_2)
		\rangle_\ast \supseteq (2\pi)^3 \delta(\vect{k}_1 +
		\vect{k}_2) \frac{P_\ast(k)^2}{\pi^2}
		\left( \frac{1}{48} k^3 \ln k - \frac{1}{90} k^3
		+ \cdots \right)
		\label{eq:twopoint-seven} .
	\end{equation}

	\subsection{Infra-red behaviour}
	
	Having obtained the relevant contributions to the two-point function,
	given by Eqs.~\eref{eq:twopoint-one},
	\eref{eq:twopoint-two},
	\eref{eq:twopoint-three},
	\eref{eq:twopoint-four},
	\eref{eq:twopoint-five},
	\eref{eq:twopoint-six}
	and~\eref{eq:twopoint-seven},
	we may collect these quantities to obtain an estimate of the total
	loop correction. It can be written
	\begin{equation}
		\langle \delta\phi(\vect{k}_1) \delta\phi(\vect{k}_2) \rangle_\ast
		\sim (2\pi)^3 \delta(\vect{k}_1 + \vect{k}_2) P_\ast(k) \Ps_\ast
		\left( -\frac{4}{3} \ln k + \beta \right) ,
		\label{eq:twopoint-total}
	\end{equation}
	where $\beta$ is an unknown renormalization-scheme dependent
	quantity left over after cancellation of the ultraviolet divergences;
	it implicitly contains the
	(ultraviolet) scale which makes $\ln k$ dimensionless.
	The coefficient of the logarithm, however, is scheme-independent
	\cite{Weinberg:2005vy,Weinberg:1995mt}.
	
	\section{Discussion}
	\label{sec:discussion}
	
	In this paper, I have computed estimates for the
	leading radiative corrections to the one- and two-point
	expectation values of the inflaton field perturbation during a phase
	of single-field, slow-roll inflation.
	After suitable ultraviolet renormalization, the loop correction
	to the one-point function
	was found to be given
	by an arbitrary renormalization-scheme dependent
	constant which can be absorbed into the background value of
	$\phi$.
	On the other hand, the loop correction to the
	two point function yielded a correction to the power spectrum of the form
	\begin{equation}
		\Ps_\ast^{\lp} =
		\Ps_\ast\Big(
			1 - \frac{4}{3} \Ps_\ast \ln k + \cdots
		\Big) .
		\label{eq:total-loop}
	\end{equation}
	
	Although the amplitude of the $\delta\phi$ power spectrum
	itself is not observable, the amplitude of $\zeta$ is accurately known
	to be of order $10^{-10}$. At tree-level in a single-field
	model, the two are approximately related
	via $\Ps_\zeta \sim \Ps_\ast/\epsilon$,
	where $\epsilon$ is the slow-roll parameter introduced in
	Eq.~\eref{eq:epsilon-def}. Since $\epsilon$ is expected to be of
	order $10^{-2}$ or less, we can conservatively suppose
	that $\Ps_\ast \lesssim 10^{-10}$. The loop correction
	given by Eq.~\eref{eq:total-loop} is therefore extremely small
	provided that the logarithm is not too large.
	
	This does not yet allow us to conclude that loop corrections are too
	small to be observable in the CMB, because it is the loop
	corrections in $\zeta$ rather than the $\delta\phi$ themselves which
	are accessible to experiment. Therefore, the prediction~\eref{eq:total-loop}
	must be translated into a prediction for
	$\Ps_\zeta^{\lp}$ before a final determination concerning the magnitude
	of loop corrections can be made. This calculation will be
	presented elsewhere \cite{Seery:2007wf}.
	However, it is already clear from
	Eq.~\eref{eq:total-loop} that \emph{quantum} effects do not greatly disturb
	the magnitude of the fluctuations imprinted in $\delta\phi$ as
	successive $\vect{k}$-modes pass outside the horizon. It is only
	the accumulation
	of fluctuations on superhorizon scales, where the fields are in an
	effectively classical regime, which can give rise to a large loop
	correction.
	
	Eq.~\eref{eq:total-loop} is
	consistent with previous estimates which
	have been made in the literature. In particular, Weinberg has estimated
	a correction to $\Ps_\ast$ from matter loops in a multi-field theory
	\cite{Weinberg:2005vy}
	which has the same functional form as~\eref{eq:total-loop}.
	Sloth \cite{Sloth:2006az,Sloth:2006nu} has given a similar estimate,
	based on the same action given in Eqs.~\eref{eq:quadratic}--%
	\eref{eq:quartic}, but evaluated several tens of e-folds after horizon
	crossing when large infrared divergences can compensate for a suppression
	in powers of slow-roll parameters; in this limit a different set of
	terms extracted from Eq.~\eref{eq:quartic} dominate the
	loop correction. This loop correction is proportional
	to $\langle \delta\phi^2 \rangle \sim \Ps_\ast \ln(k)$
	for a flat spectrum, which reproduces the logarithmic $k$-dependence
	described by~\eref{eq:total-loop}.
		
	\ack
	I acknowledge support from PPARC under grant PPA/G/S/2003/00076.
	I would like to thank M. Sloth, D. Lyth, K. Malik, J. Lidsey,
	C. Byrnes and A. Mazumdar
	for useful conversations, and especially F. Vernizzi
	for lengthy conversations and correspondence which have helped
	clarify my understanding. I would like to acknowledge the hospitality
	of the Abdus Salam Institute for Theoretical Physics, Trieste,
	and the Department of Physics, University of Cardiff, where portions
	of the work outlined in this paper were carried out.
	
	I would like to thank P. Adshead, R. Easther and E. Lim for drawing
	my attention to a sign error in the Feynman rules of
	{\S}\ref{sec:quantum-rules} which appeared in earlier
	versions of this paper.
	
	\section*{References}

\providecommand{\href}[2]{#2}\begingroup\raggedright\endgroup

\end{fmffile}
\end{document}